\documentclass[letterpaper]{article}
\usepackage[T1]{fontenc}
\usepackage{aaai}
\usepackage{times}
\usepackage{helvet}
\usepackage{courier}
\usepackage{subfigure}
\usepackage{commath}
\usepackage{graphicx}
\usepackage{url}
\usepackage{epstopdf}
\usepackage{fancyhdr}
\usepackage{rotating}
\usepackage{subfigure}
\usepackage{amssymb}
\usepackage{amsmath}
\usepackage{multirow}
\usepackage{multicol}
\usepackage{indentfirst}
\usepackage{float}
\usepackage{epsfig}
\usepackage{algorithm,algorithmic}
\usepackage{color}

\newcommand{\remove}[1]{}

\begin{document}

\title{Network Weirdness: Exploring the Origins of Network Paradoxes}
\author{
Farshad Kooti \\
USC Information Sciences Institute\\
4676 Admiralty Way \\
Marina del Rey, CA 90292 \\
\texttt{kooti@usc.edu} \\
\And
Nathan O. Hodas \\
USC Information Sciences Institute\\
4676 Admiralty Way \\
Marina del Rey, CA 90292 \\
\texttt{nhodas@isi.edu} \\
\And
Kristina Lerman \\
USC Information Sciences Institute\\
4676 Admiralty Way \\
Marina del Rey, CA 90292 \\
\texttt{lerman@isi.edu}
}
\maketitle

\begin{abstract}
Social networks have many counter-intuitive properties, including the ``friendship paradox''  that states, on average, your friends have more friends than you do. Recently, a variety of other paradoxes were demonstrated in online social networks. This paper explores the origins of these network paradoxes. Specifically, we ask whether they arise from mathematical properties of the networks or whether they have a behavioral origin.
We show that sampling from heavy-tailed distributions always gives rise to a paradox in the mean, but not the median.
%We propose a stronger form of network paradoxes, based on utilizing the median, whereby the majority of your friends and followers have more friends, followers, activity, etc. than you do. We use data from two online social networks to explore the origins of the strong paradoxes on networks. First, we demonstrate that network paradoxes exist for the median, ruling out a purely mathematical explanation.
We propose a strong form of network paradoxes, based on utilizing the median, and validate it empirically using data from two online social networks. Specifically, we show that for any user the majority of user's friends and followers have more friends, followers, etc. than the user, and that this cannot be explained by statistical properties of sampling.
%Thus, the magnitude of the strong paradox can be used to reveal behavioral factors leading to these paradoxes.
Next, we explore the behavioral origins of the paradoxes by using the shuffle test to remove correlations between node degrees and attributes.  We find that paradoxes for the mean persist in the shuffled network, but not for the median. We demonstrate that strong paradoxes arise due to the assortativity of user attributes, including degree, and correlation between degree and attribute.

%Social networks have many counter-intuitive properties, most famously, the ``friendship paradox.'' Recently, a variety of other paradoxes were demonstrated. This paper explores the origins of these paradoxes. Specifically, we ask whether they only arise from mathematical properties of the networks or whether they are behavioral in origin. We show that sampling from heavy-tailed distributions always gives rise to a paradox in the mean, but not the median in the shuffled network. Therefore, we show that median is a more robust measure of the paradox.
%
%We use data from two social networks Twitter and Digg to empirically verify our findings. We conduct shuffle tests and show that paradoxes still exist for mean but not for median. We demonstrate that the paradoxes are due to some behavioral processes of network formation and maintenance and not simply the statistical properties of the distribution. We also show that the paradox in the characteristics of the users is mostly due to the assortativity in that characteristic but assortativity can not fully account for observed paradoxes.

\end{abstract}

\section{Introduction}

The interplay between individual's attributes and her choice of who to link to in a social network produces much of the network's observed complexity and leads to counter-intuitive phenomena, such as the ``friendship paradox''. This paradox states that regardless of which individual you pick in a social network, on average, her friends have more friends than she does~\cite{Feld91}. The paradox has been observed both in online~\cite{Ugander11,Hodas13icwsm} and offline social networks~\cite{Feld91,zuckerman2001makes,Eom14}. Paradoxes for attributes other than the number of friends have also been observed. For example, a Twitter user posts fewer messages and receives less-viral content than her friends do on average~\cite{Hodas13icwsm}. Similarly, a scientist's co-authors are more productive and better cited on average~\cite{Eom14}. Additionally, paradoxes can lead to systematic biases in how individuals perceive their world~\cite{zuckerman2001makes,sgourev2006lake,wolfson2000students,yoganarasimhan2012impact,kanai2012brain}
and in how peers affect an individual's behavior. For instance, studies have shown that teenagers over-estimate alcohol and drug use of their peers, because the correlation between connectivity and drug/alcohol use allows popular drinkers and smokers to skew perceptions of their many friends~\cite{tucker2011substance,wolfson2000students}. On a positive note, the friendship paradox has been used as a basis for early detection of flu outbreaks on a college campus~\cite{Christakis10} and of trending topics on social media~\cite{garcia2012using}.

We examine the origins of these network paradoxes. Specifically, we ask whether the paradoxes are simply a consequence of the mathematical properties of networks and populations of users comprising them or whether they have a behavioral origin.
First, we show that the conventional measure of paradox, which compares a user's attribute, e.g., number of neighbors, to the \emph{mean} of the attributes of her network neighbors, will almost always produce a paradox, even if the network is completely random.   This can be explained by the statistical properties of sampling from a heavy-tailed distribution, and many social attributes, such as degree, have such heavy-tailed distributions. Hence, it is not surprising to observe this sort of `traditional paradox' in the mean on any social network for almost any attribute.  We show that using the median, instead of the mean, is a more robust measure of ``paradoxical behavior''.

Next, we measure network paradoxes on social media sites Digg and Twitter. We confirm traditional paradoxes for several user attributes, including number of neighbors, activity, and also diversity and virality of content user sees. In addition,  we find that paradoxes persist when users' attributes are compared to the \emph{median}, rather than the mean, of the attributes of their neighbors. As a consequence, network paradoxes can be restated in a stronger form:
For a variety of attributes, a majority of a user's neighbors have a value that exceeds the user's own value, i.e., not only do user's friends have more friends on average, but \emph{a majority of friends have more friends than the user}.
Because the strong paradox does not trivially arise from the statistical properties of distributions, it must have a behavioral origin.

To better understand behavioral causes of network paradoxes, we perform  shuffle tests to destroy correlations between degree and user attributes. There exist two types of correlations in social networks. First, there is a correlation between attribute of a node, such as activity, and its degree (within-node correlation). In addition, there is assortativity, or correlation between the attributes of a node and the attributes of its neighbors (between-node correlation). Each of these correlations might explain the paradoxes. To understand the exact origin, we separate the effect of different correlations by destroying one of the correlations while keeping the other one, which is done using a controlled shuffle test. We find that  both within-node and between-node correlations are each responsible for the paradoxes.

%In short, our work characterizes the differences between the traditional (`weak') paradox, which is calculated using the mean, and our strong paradox, which is calculated using the median.  We empirically demonstrate that previously observed paradoxes still hold as strong paradoxes.  However, upon shuffling the network, weak paradoxes persist but the strong paradox is destroyed or diminished, demonstrating that the magnitude of strong paradox can be used as a measure of behavioral factors in the creation and maintenance of the social network.  We also show that the observation of a paradox in the mean does not imply any inter-node or intra-node correlation between attributes and connectivity; they can arise in completely random networks.   Our work suggests, for Twitter and Digg, network structure and node attributes are intimately connected. Accounting for these often non-intuitive relationships is necessary for understanding, modeling, and predicting individual and social behavior in networks.

%the feedback between structure and attributes affects both individual's behavior and evolution of social networks~\cite{Kossinets:2009}. Understanding this feedback is key to describing,

Existence of the strong paradox for an attribute implies that for most users, a randomly selected friend is likely to exceed the user in that attribute. Because the same behavioral factors that  create between-node correlations in attributes and within-node correlations  are often related to desirability of that attribute (extraversion, wealth, etc.), people end up dynamically positioning themselves in the network to remain subject to the strong paradox. The users will perceive themselves as inferior to friends, even if comparison is done on a one-to-one basis. This may explain why self-assessments are negatively correlated with exposure to online social media~\cite{kross2013facebook,chou2012they}.

Our specific contributions are as follows:
\begin{enumerate}
\item We characterize the differences between the traditional (`weak') paradox, which is based on the mean, and our strong paradox, based on the median.  We empirically demonstrate that previously observed paradoxes still hold as strong paradoxes.

\item We show that upon shuffling the network, weak paradoxes persist but the strong paradoxes are diminished or destroyed, demonstrating that the magnitude of strong paradox can be used as a measure of behavioral factors in the creation and maintenance of the social network.

\item We also show that the observation of a weak paradox does not imply any within-node or between-node correlation between attributes and connectivity; they can arise in completely random networks devoid of correlation.
\end{enumerate}

Our work suggests that  network structure and node attributes in social media are intimately connected. Accounting for these often non-intuitive relationships is necessary for understanding, modeling, and predicting individual and social behavior in networks.

\if 0
 A variety of paradoxes have been observed in social networks, including the friendship paradox~\cite{Feld91,Ugander11}, activity and virality paradoxes~\cite{Hodas13icwsm}.
 The friendship paradox states ...
 These paradoxes have counter-intuitive implications.

 Homophily is an often-observed consequence: similar nodes are more likely to be linked in social networks.

 \begin{itemize}
 \item friendship paradox was used as a basis for early detection of influenza outbreaks~\cite{Christakis10}

 \item trending topics on Twitter~\cite{garcia2012using}

 \item information overload and consequences~\cite{Hodas13icwsm}

 \item may cause systematic biases in the estimates people make of the global state of the system from local observations, with consequences on social perception, decisions people make, evolution of social norms, and collective action (e.g., ``do I join a protest?'').

 \item may affect how people position themselves in the network to receive high utility information~\cite{Burt95}\cite{Burt04}\cite{Aral11}
 \end{itemize}

First, we show that sampling from a heavy-tailed distribution leads to a paradox in the mean, but not the median. Median is, therefore, a more robust measure for the existence of a paradox.

 Then, we perform a shuffle test to examine behavioral causes:

- correlation of connectivity (degree assortativity)

- correlation of activity (activity assortativity)

- correlation between activity and degree

\fi

\section{Statistical Origins of Paradoxes}
\label{sec:statistical}

\begin{figure}[thb!]
%\begin{center}
\begin{tabular}{@{}c@{}c@{}}
\subfigure[ ]{
   \includegraphics[width=0.5\columnwidth]{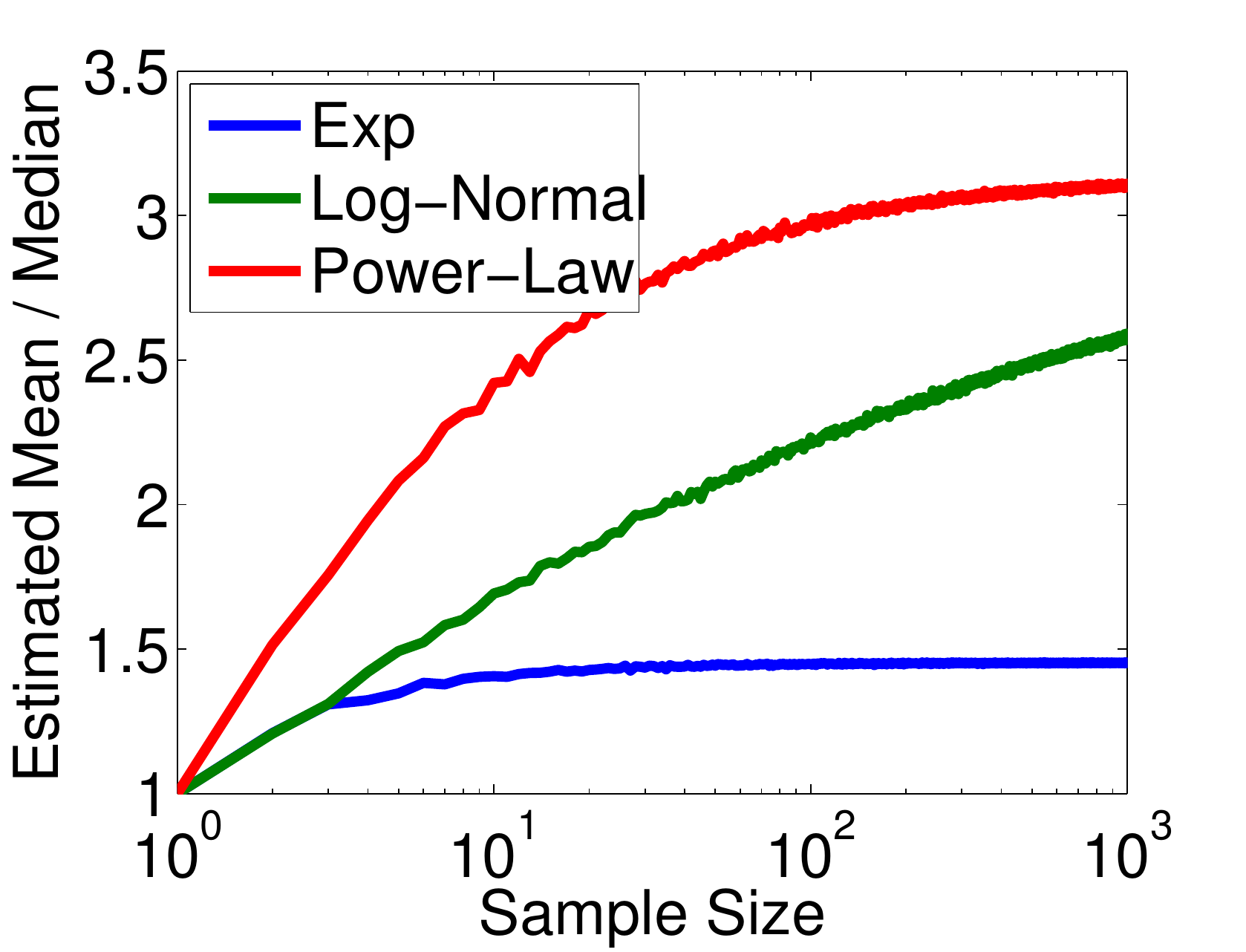}
      \label{fig:scaling}
   }% requires the graphicx package
  &
   \subfigure[ ]{
   \includegraphics[width=0.5\columnwidth]{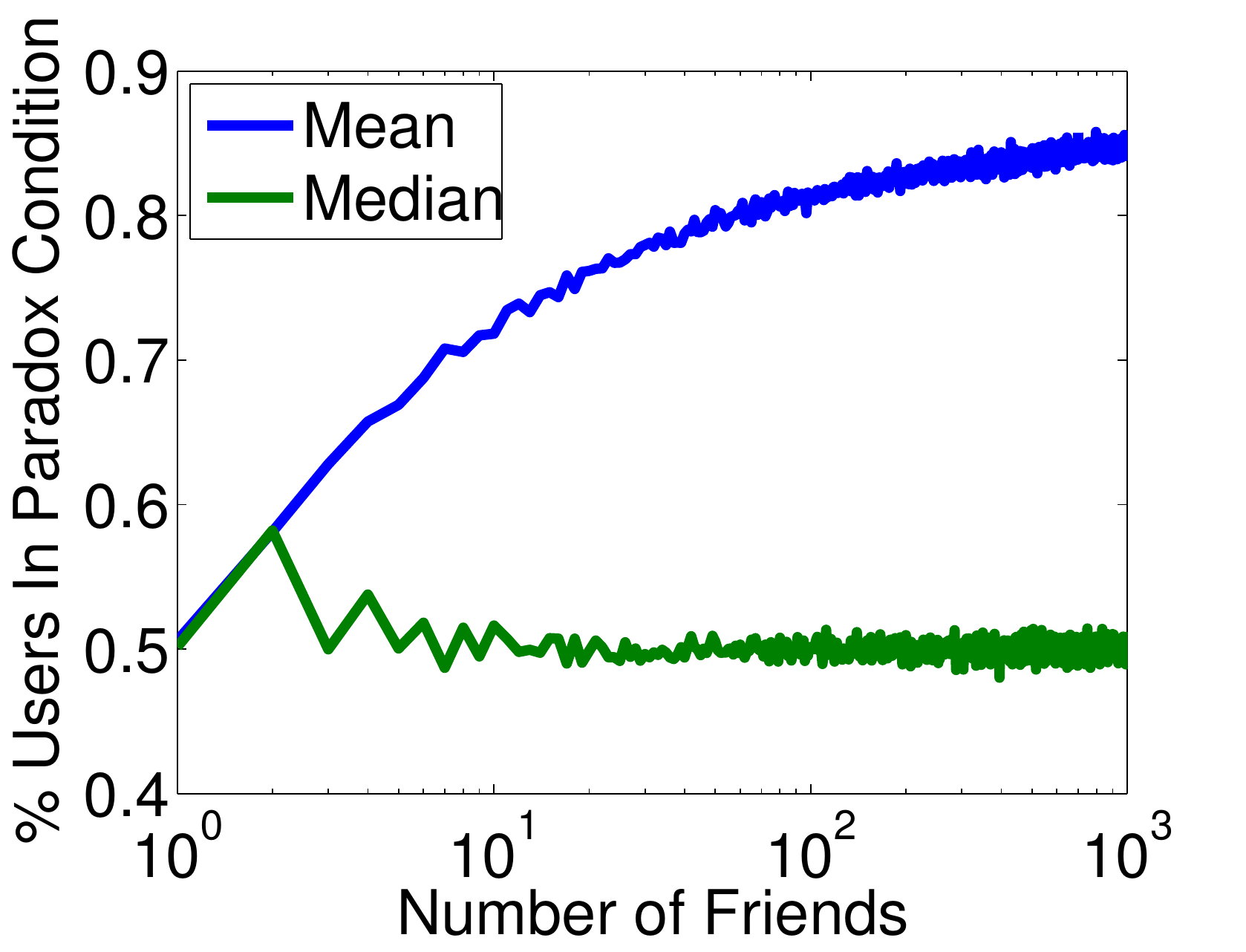}
   \label{fig:meanvmedian}
   }
  \end{tabular}
   \caption{(a) Estimated mean grows with sample size.  Three distributions, exponential (Exp), log-normal, and Pareto, each result in different biased estimates of the mean.  The larger the difference between median and population mean, the larger the discrepancy. Thus, a user will always observe a paradox when calculating the mean of its neighbors when the population mean is greater than the median. (b) Effect of using mean vs. median on fraction of users with given number of friends estimated to be in paradox condition in a random network with no correlations. Users' attributes are drawn independently from $\sim x^{-1.2}$.}
 %  \label{fig:scaling}
  %\end{center}
\end{figure}

The friendship paradox is thought to be rooted in the heterogeneous distribution of node attributes, such as node degree~\cite{Feld91}. Such distributions are characterized by a ``heavy tail'', where extremely large values, e.g., high degree nodes, appear much more frequently than expected compared to a normal or exponential distribution. These large values skew the ``average'', giving rise to a large difference between the mean and the median. %It is widely accepted that the median of a long-tailed distribution is less sensitive to extreme value occurrences than the mean, and is a more robust measure of the ``average''. NOH -- true, but I think in the present case, we are purely interested in how one the estimates of mean and median vary with sample size.
In this section, we show that randomly sampling from a heavy-tailed distribution can produce a paradox when using the mean, but not the median. Therefore, if the friendship paradox also exists in the median, it cannot be purely statistical in nature, and must have a behavioral origin.  Conversely, a paradox using the mean may arise simply from statistics of the attribute distribution, without any behavioral component or correlation between users.%The nature of the present paradoxes depend on how one defines ``average."  Because of the long-tail of distributions such as degree, activity, and vitality, a large difference between median and mean may exist.  As we will show, sampling from Pareto distribution leads to paradoxes in the mean, but not the median, of many of the paradoxes we investigate.

To be more precise, consider the definition of mean vs median for continuous, non-negative distributions\footnote{The conclusions in the present work hold for any distribution where the mean is greater than the median.  This is almost always the case for heavy-tailed distributions, but may be violated for small-support discrete distributions.  When median > mean, the present conclusions are simply reversed; we would expect `anti-paradoxes' when considering the mean.}. The mean is defined as $\mu \equiv \int_0^\infty x P(x)\,dx.$
\remove{KL: should we focus on discrete distributions instead?  Because we are focusing on generic features of distributions, we should stay in the continuous regime}
The median, $m$, is defined by the solution to $\frac{1}{2} = \int_0^m P(x)\,dx$.  Given a sample consisting of a single random instance drawn from $P(x)$, there is an equal chance that it will be larger or smaller than $m$.  What value would you estimate that minimizes the mean absolute error with the mean of a sample with size = 1? This is the median of the distribution, $m$~\cite{lee1995graphical}. At the other extreme, the value that minimizes the absolute error with the mean of an infinitely large sample is $\mu$, by definition.
Social behavior is often characterized by unimodal distributions with heavy tails, so $m \leq \mu$.  Thus, as the sample size from a distribution increases, the observed mean of the sample increases monotonically from $m$ to $\mu$.

\remove{
\begin{figure}[htb]
\begin{center}
   \includegraphics[width=0.9\columnwidth]{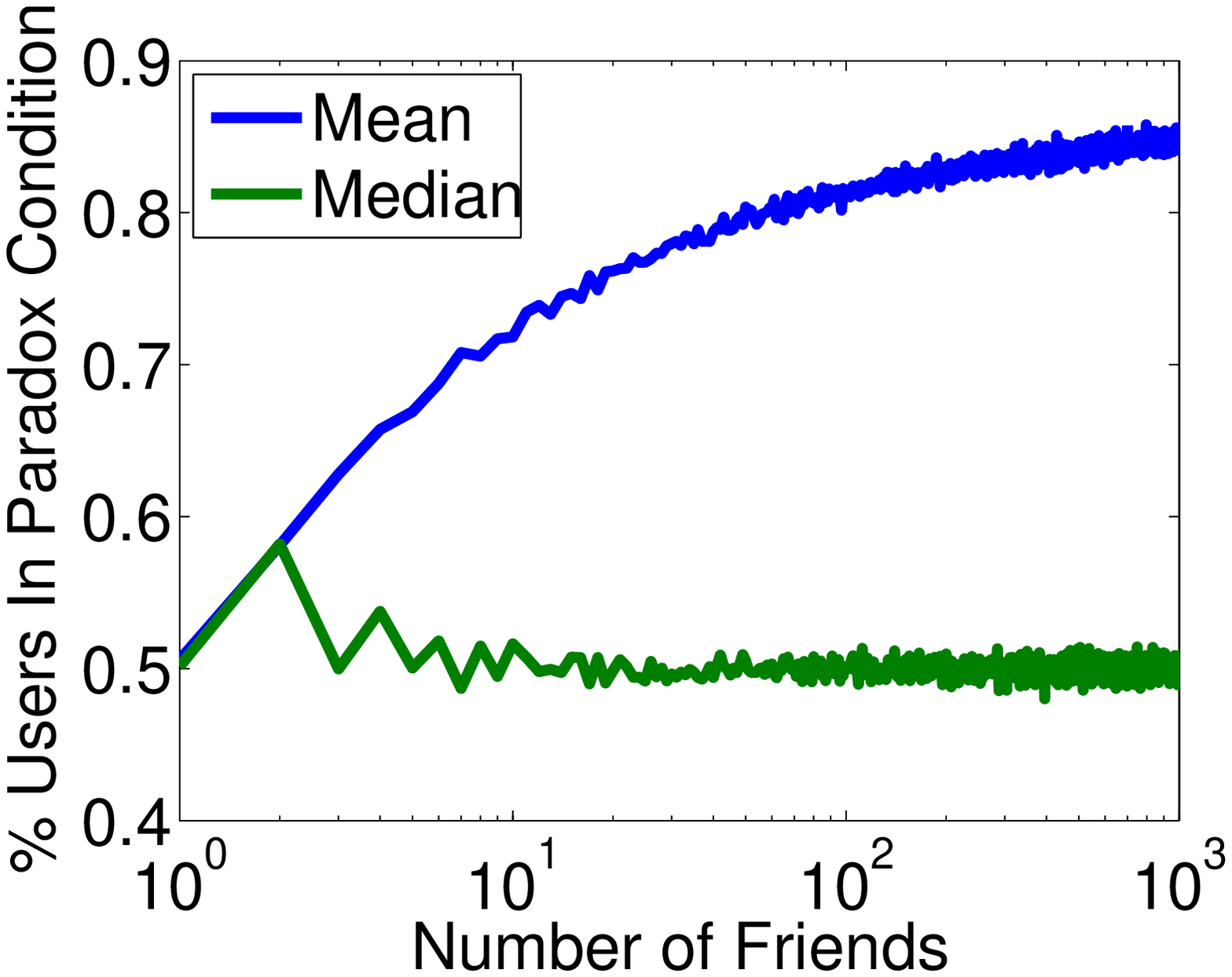} % requires the graphicx package
   \caption{Effect of using mean vs. median on fraction of users with given number of friends estimated to be in paradox condition in a random network with no correlations. Users' attributes are drawn independently from $\sim x^{-1.2}$.}
   \label{fig:meanvmedian}
   \end{center}
\end{figure}
}

Figure~\ref{fig:scaling} provides an illustration of this behavior. We randomly sample values from three example distributions, exponential ($\sim e^{-2x}$), log-normal ($\mu=-0.3,\sigma=1.5$), and Pareto ($\sim x^{-1.2}$), and calculate the mean and median of
the samples of varying sizes.  The true mean and median for these distributions are $(\mu=\frac{1}{2}, m=\frac{1}{2}\log 2)$, $(\mu=2.28,m=0.74)$, and $(\mu=6,m=1.78)$,   respectively. As explained  above, the estimated sample mean changes monotonically from $m$ to $\mu$ as the sample size increases from 1, shown in Fig.~\ref{fig:scaling}. However, the sample median does not vary with sample size, because half of the numbers in the sample are below the population median and half above.  Thus, if you consider the fraction of users in paradox condition, shown in Figure~\ref{fig:meanvmedian}, when users attributes are drawn iid from the previous Pareto distribution, the more friends a user has, the more likely the mean of their friends' attributes exceeds their own, but when using the median, no paradox is observed.

Thus, consider the following explanation of how network paradoxes arise and how they depend on what you mean by ``average.''  Assume we are measuring  some empirical quantity --- user attribute $x$ --- in a purely random network where each user's $x$ is an independent, identically distributed (iid) variable (and $x$ is not node degree).  The $x$ for a user is compared to the ``average'' $x$ for the user's network neighbors.  The best estimate (with respect to mean absolute error) of $x$ for the user is $m_x$, because it is a sample of size 1.  On the other hand, the number of users' neighbors is at least one, meaning that the best estimate for the mean of the neighbors'  $x$ is $\geq m_x$.  Therefore, even in a purely random network, as long as the mean of $x$ is greater than the median of $x$, one would be led to the conclusion that ``\remove{in the mean, }your $x$ is smaller than the mean of your neighbors' $x$.'' But, if you consider the \emph{median}, both you and your neighbors will have the same median, and no paradox will be observed.  One may show that in a fully-connected network, where attributes are iid, the fraction of users in the strong paradox condition is no larger  than $0.5 + 1/N$, but the weak paradox condition may hold for as many as $N-1$ of the users. 

Observation of paradoxes utilizing the median provides a test of the origin of such paradoxes: do they arise simply from heavy-tailed distributions or are they due to humans positioning themselves in the network according to some nontrivial behavioral mechanism?

\remove{Consider a series of random variables $x_i$. The probability that $x_0 < (x_1+x_2)/2$ is
\remove{\begin{align}
 =&\int_0^\infty \, dx_0 P(x_0) \int_0^{2x_0}\,dx_1 P(x_1) \int_0^{2(x_0 - x_1)}\, dx_2 P(x_2)\\
	=& \iiint_0^\infty P(x_0)P(x_1)P(x_2)\theta\left((x_1+x_2)/2 - x0\right) \,dx_0dx_1dx_2.
\end{align}}}

\section{Measuring Network Paradoxes}
\label{sec:empirical}

In the previous section, we showed that a network paradox could always exist when considering
the mean of some user attribute but not necessarily the median. In this section, we analyze online social networks of Twitter and Digg users and show that network paradoxes exist for several user attributes when considering both the mean and median. Therefore, they cannot arise simply due to properties of heavy-tailed distribution of user attributes.

\subsection{Data}

We use the Twitter dataset gathered by~\cite{Yang2011wsdm}. The dataset includes
476M tweets, which are 20-30\% of all tweets posted between June and December 2009.
We complement this dataset with the social network of Twitter as of August 2009~\cite{Kwak10www}.
To guarantee that we have information about all user's connections and activity data, we only use the first two months of tweets, because we do not have the follow links that were established later.
We also only consider the links that are between users tweeted at least once during the observation period. The remaining network includes 5.8M users with 193.9M follow links. This network was used to measure friendship paradoxes.

%For activity and virality paradoxes, we further shrink the dataset to have a fair comparison:
For measuring other network paradoxes, we further restrict the network to users who joined Twitter before the beginning of the dataset. These are the users whose activity is recorded over the entire observation period. This leaves us with 29.4M tweets, with 2.2M users and 113M links.

In addition, we also use the Digg dataset presented by~\cite{sharara:icwsm11}.
Digg is distinct from Twitter because it is primarily a social news site. Despite this, it is similar to Twitter in many ways. Digg users submit links to news stories they find online, which is similar to posting a tweet. Other users vote for, or digg, these stories, which is similar to retweeting. Moreover, Digg allowed users to follow submissions and diggs of other users, creating a friend-follower network similar to Twitter. However, unlike Twitter, Digg promoted popular news stories to the front page, where they could be seen by all Digg users. To reduce potentially confounding factors, we focus only on pre-promotion votes, when the stories were mainly visible through the social network~\cite{hodas2013attention}.

The Digg dataset includes all diggs on news stories of 11.9K users submitted to Digg over a six months period of July to December 2010. There are more than 1.9M diggs on 29K stories in the dataset, with 1.3M follow links.

For convenience, we refer to the people followed by a user $U$ as $U$'s \emph{friends} and those who follow her as $U$'s \emph{followers}. Thus, $U$ receives content from her friends in the form of messages on Twitter or recommendations for news stories on Digg,  and sends content to her followers.

\subsection{Distribution of Attributes}

\begin{figure}[t!]
\begin{center}
\begin{tabular}{cc}
\subfigure[Number of friends]{%
\label{fig:friend_pdf}
\includegraphics[width=0.45\columnwidth]{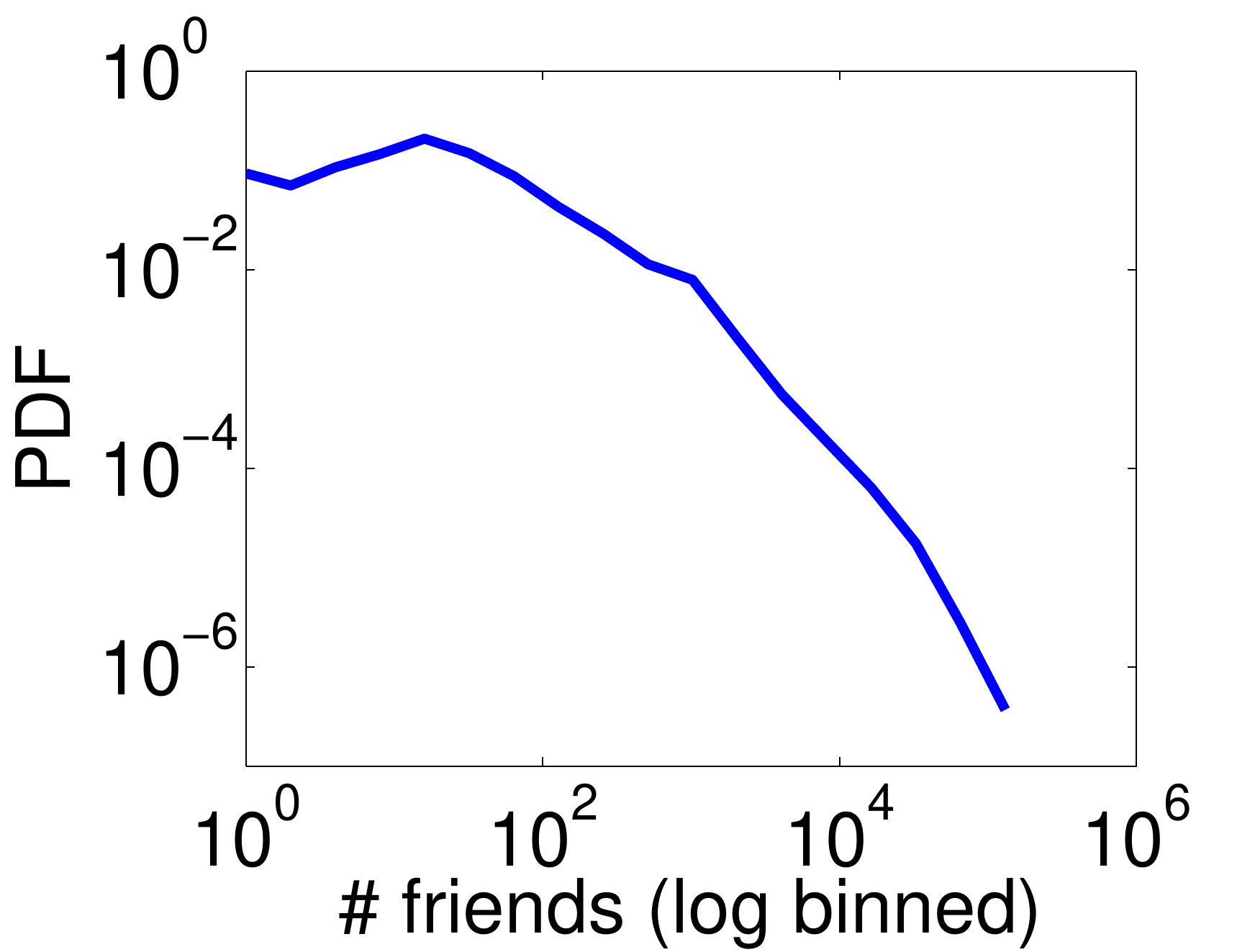}
}
&
\subfigure[Number of followers]{%
\label{fig:follower_pdf}
\includegraphics[width=0.45\columnwidth]{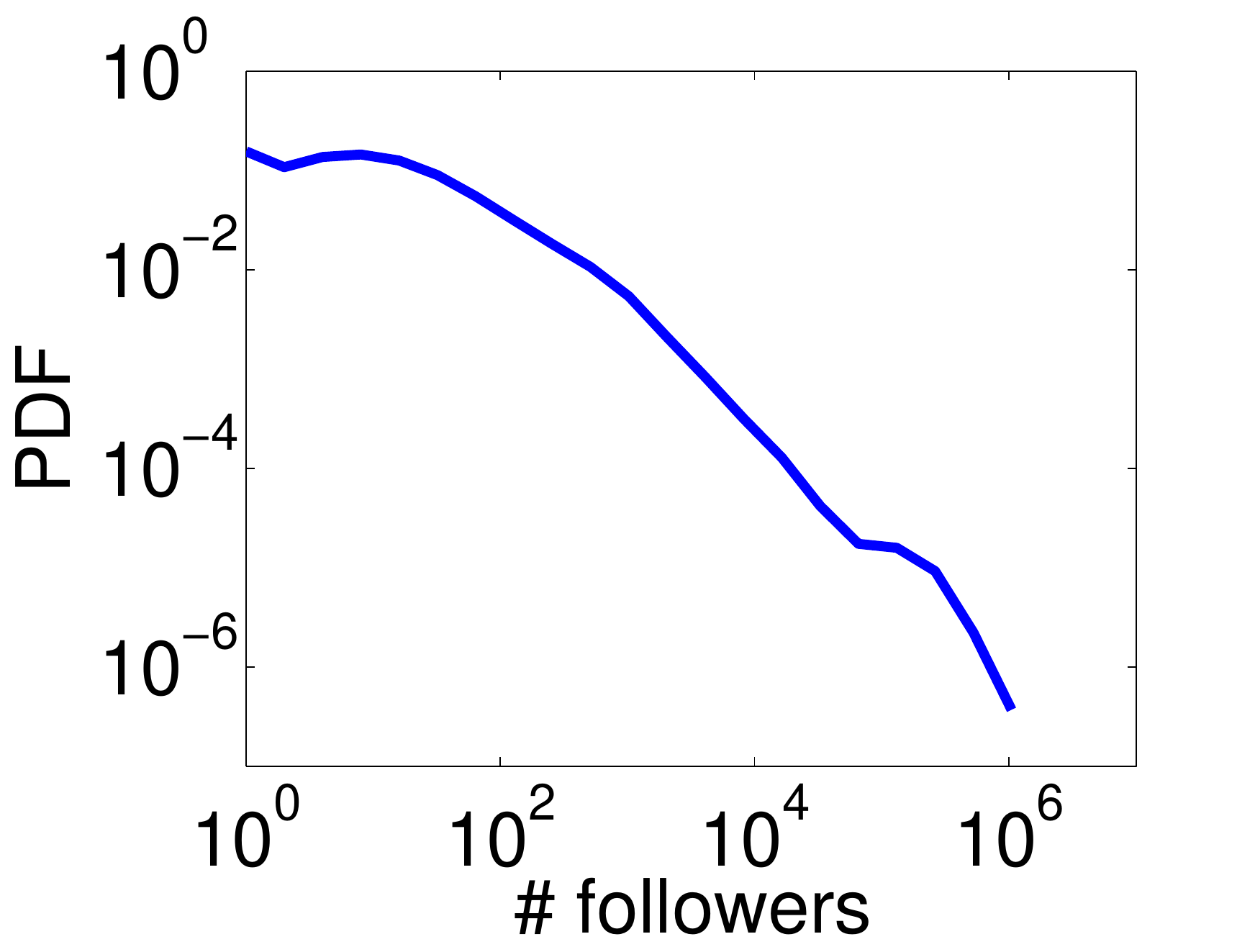}
}
\\
\subfigure[User activity]{%
\label{fig:activity_pdf}
\includegraphics[width=0.45\columnwidth]{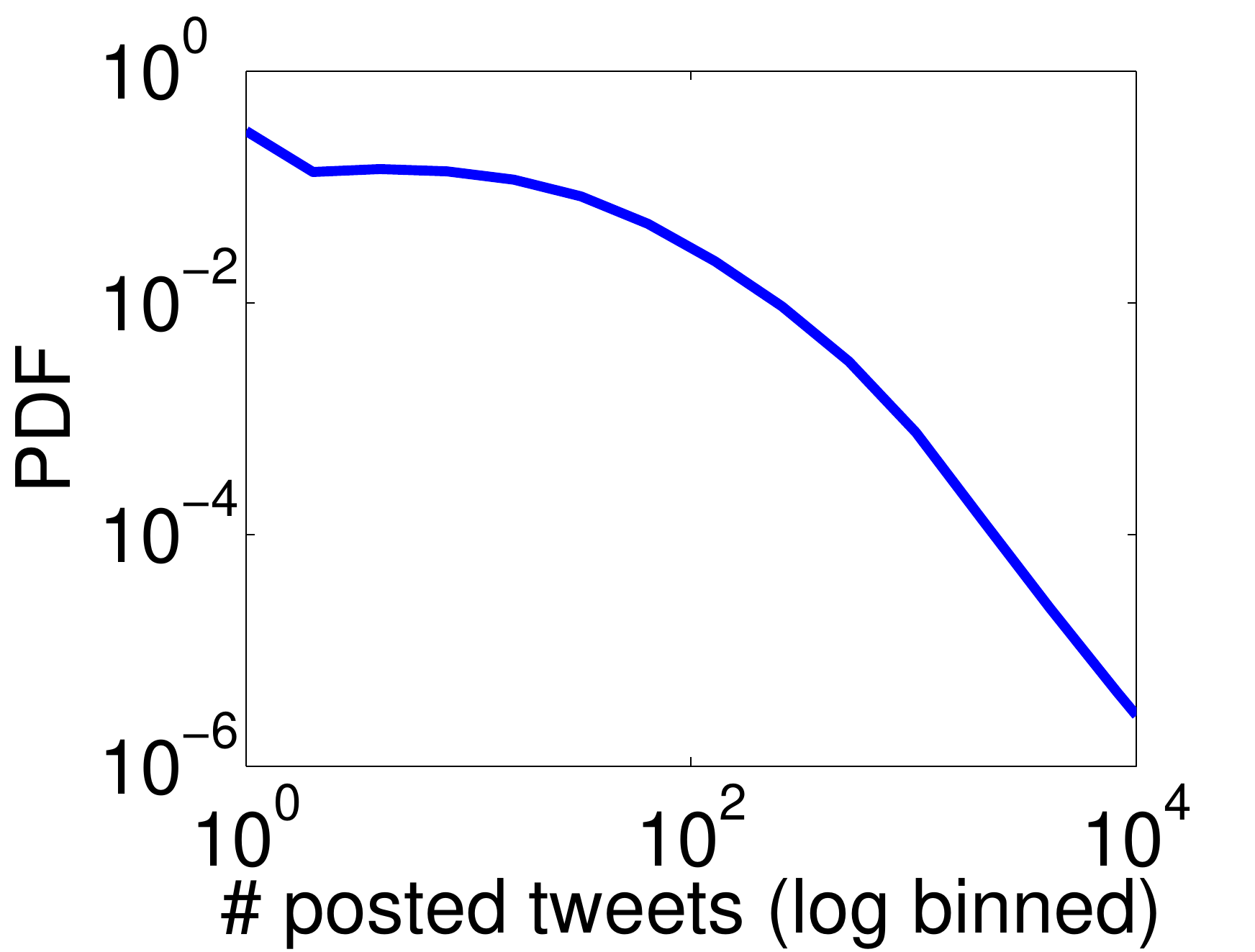}
}
&
\subfigure[Diversity of received URLs]{%
\label{fig:diversity_pdf}
\includegraphics[width=0.45\columnwidth]{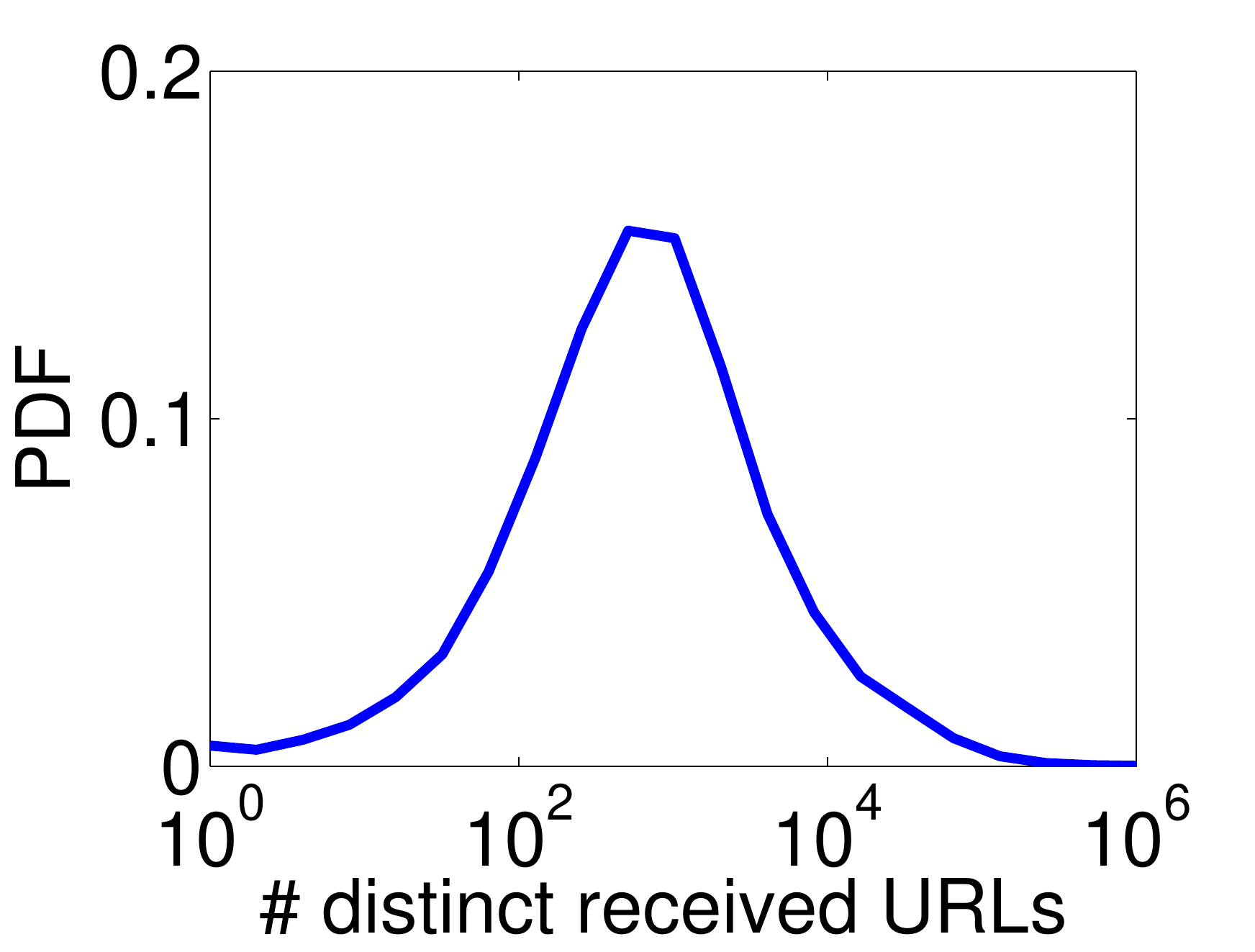}
}
\\
\subfigure[Virality of posted URLs]{%
\label{fig:virality_post_pdf}
\includegraphics[width=0.45\columnwidth]{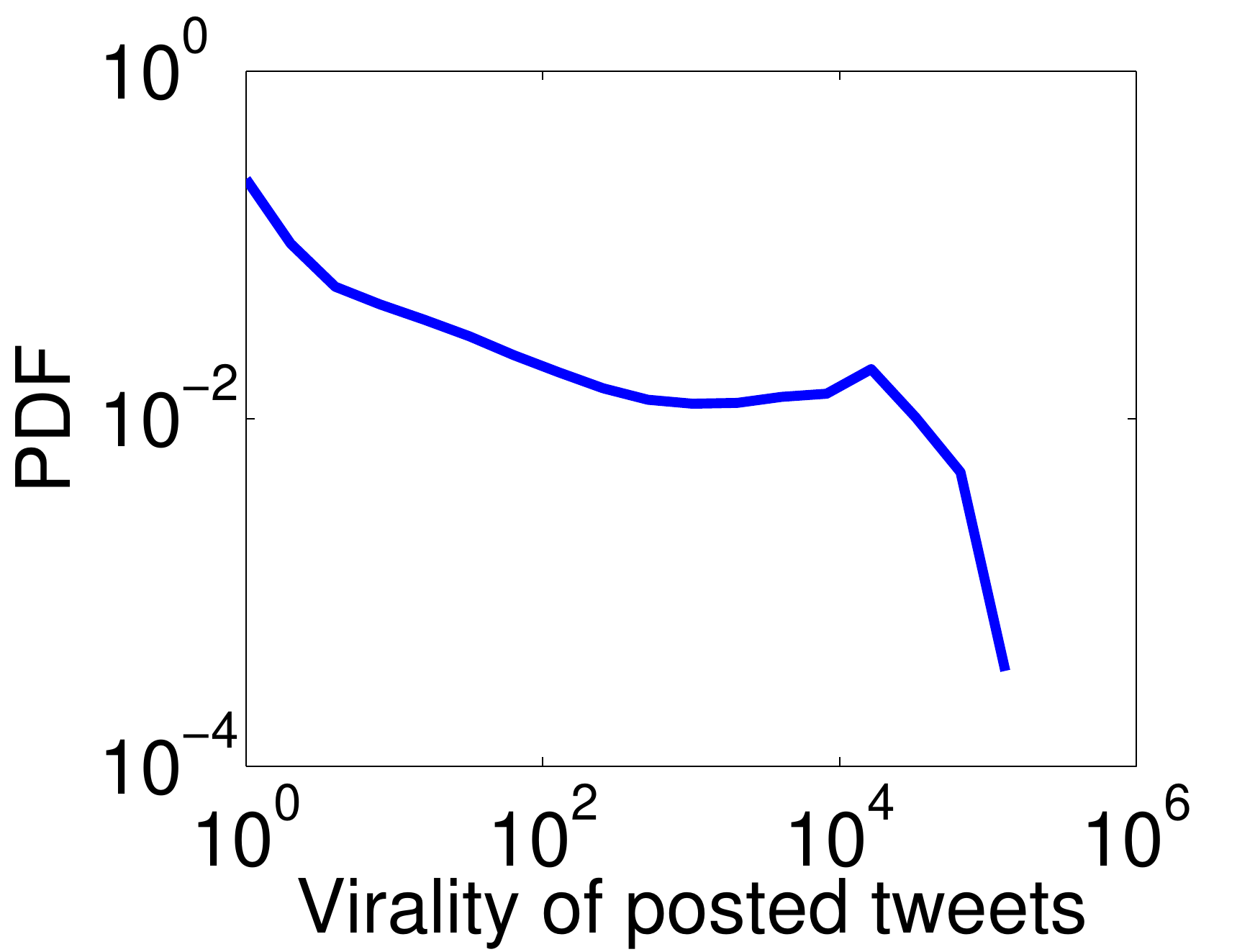}
}
&
\subfigure[Virality of received URLs]{%
\label{fig:virality_received_pdf}
\includegraphics[width=0.45\columnwidth]{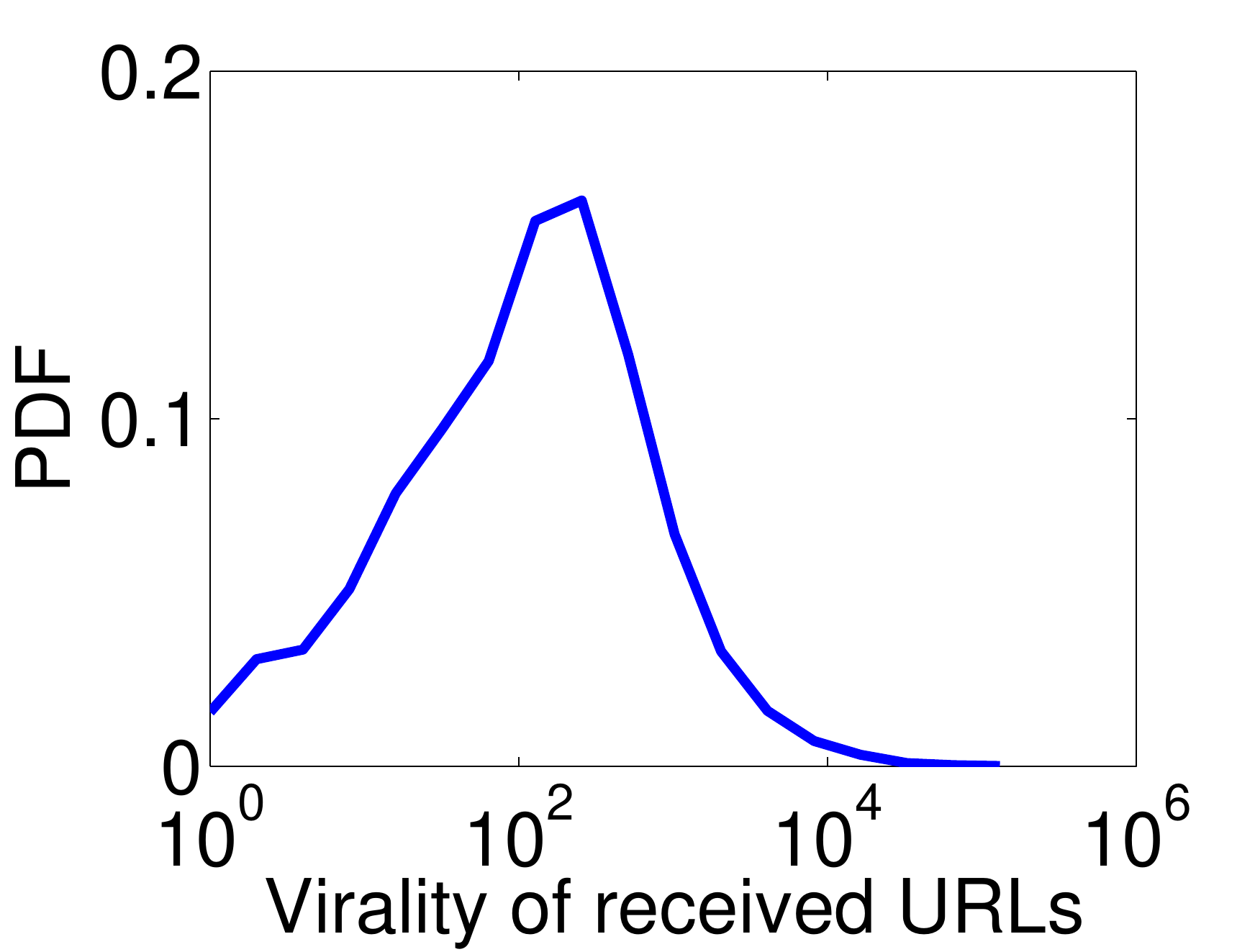}
}
\end{tabular}
\caption{Distribution of user attributes on Twitter.}
\vspace*{-2mm}
\label{fig:distributions}
\vspace*{-2mm}
\end{center}
\end{figure}

Social networks share some common structural properties, such as a heavy-tailed  degree distributions. Most of the nodes in such networks have few connections, or small degree, while few nodes have a large number of connections, or high degree.
Besides degree, many other attributes of Twitter (Digg) users have a heavy-tailed distribution. In this paper we focus on the following attributes:
\begin{description}
\item[Degree:] number of friends and followers of a user
\item[Activity:] number of tweets (diggs) made during the observation period by the user
\item[Diversity:] number of distinct URLs (or news stories on Digg) received by user from friends
\item[Virality:] popularity of content posted or received by the user, as measured by the number of retweets (or  pre-promotion votes on Digg) it receives
\end{description}
Figure~\ref{fig:distributions} shows the probability distribution of the observed values of each attribute, i.e., fraction of users in our sample who have that attribute value on Twitter. We logarithmically bin the values to reduce sparseness of extreme values. The distributions are characteristically heavy-tailed, regardless of whether the attribute depends on the decisions made by the user (number of friends, activity) or the decisions of others (number of followers, diversity). Except for diversity (Fig.~\ref{fig:diversity_pdf}) and virality of received content (Fig.~\ref{fig:virality_received_pdf}), which resemble log-normals, the user attribute distributions have a power law-like shape.
The results for Digg are similar, except for the virality of Digg posts, which is a not heavy-tailed, due to the fact that posts accumulate only a limited number of diggs before the promotion.

Besides heavy-tailed attribute distributions, networks have other important statistical regularities. For example, many social networks are assortative, meaning that nodes  tend to connect to other nodes having a similar degree~\cite{newman2002}.
%A positive degree--degree correlation is called \emph{assortativity}, and it is one of the attributes of social networks~\cite{PhysRevE.68.036122}. However, this characteristic only exists in undirected networks and not on the web~\cite{newman2002} or the Twitter follower graph~\cite{Kwak2010}. \note{NOH: If we find negative assortativity, this is good to include, as it means this generic correlation is also not the sole explanation. As shown by Eom and Jo, negative assortative should actually result in \emph{anti-paradoxes}~\cite{eom2014generalized}.}
%
In addition to degree \emph{assortativity}, other correlations may exist between attributes of connected nodes. This phenomenon, known as \emph{homophily}, is a generic property of social networks and results in connected users being similar~\cite{homophily} and becoming more similar over time~\cite{Kossinets:2009}. In addition to between-node correlations, \emph{within-node} correlations are important in social networks. The most important of these is the correlation of user's attributes, such as income, activity, or productivity, with her degree~\cite{Hodas13icwsm,Eom14}. Later in the paper we study the role of these correlations in explaining network paradoxes.

\subsection{Network Paradoxes}

\begin{figure} \begin{center}
\includegraphics[width=0.9\columnwidth]{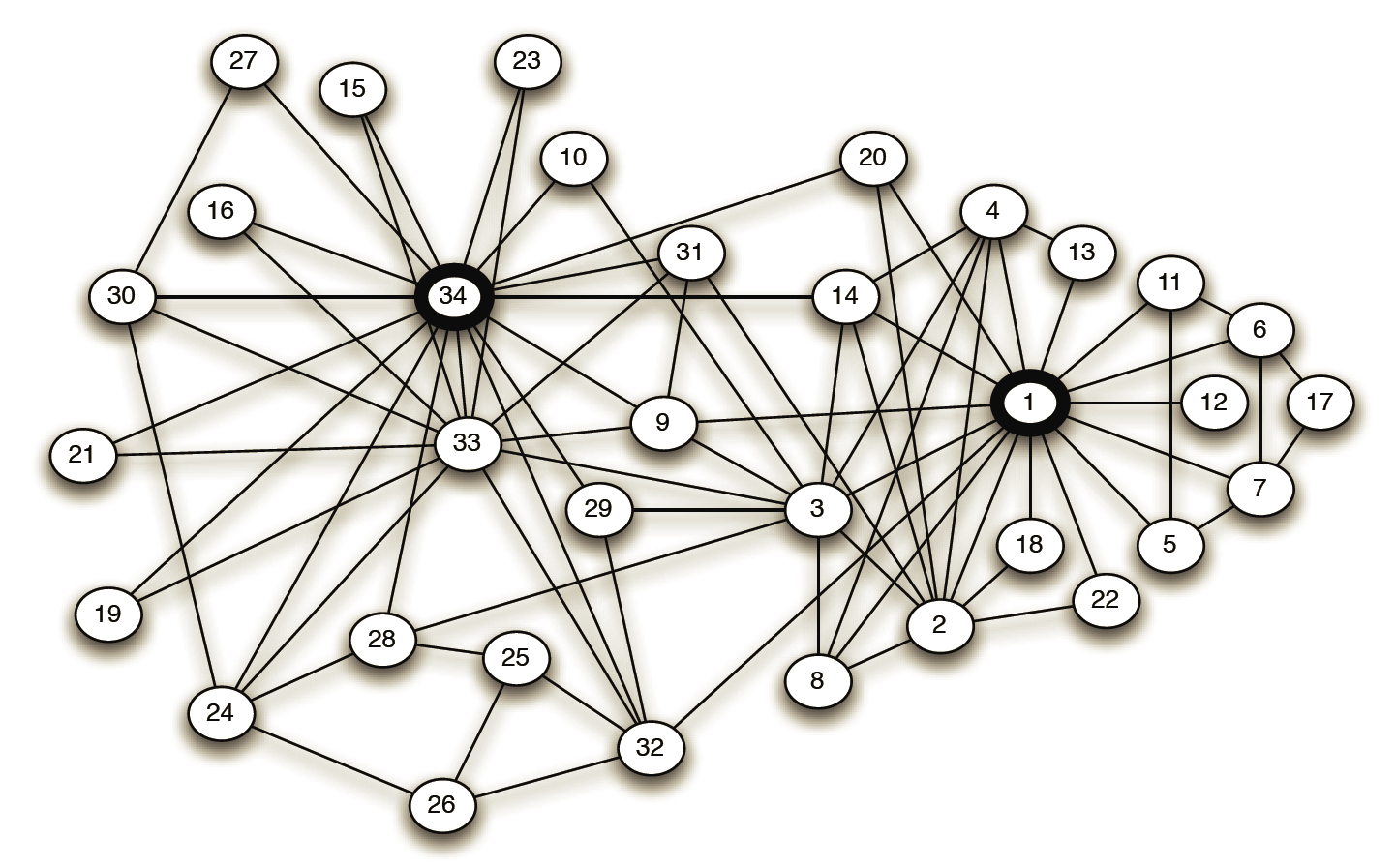}
\end{center}
\vspace*{-2mm}
\caption{Friendship network of Karate Club}
 \label{fig:karate_network}
 \vspace*{-2mm}
  \end{figure}

\begin{figure}[thb!]
\begin{tabular}{c}
\subfigure[Friendship paradox.]{
   \includegraphics[width=0.9\columnwidth]{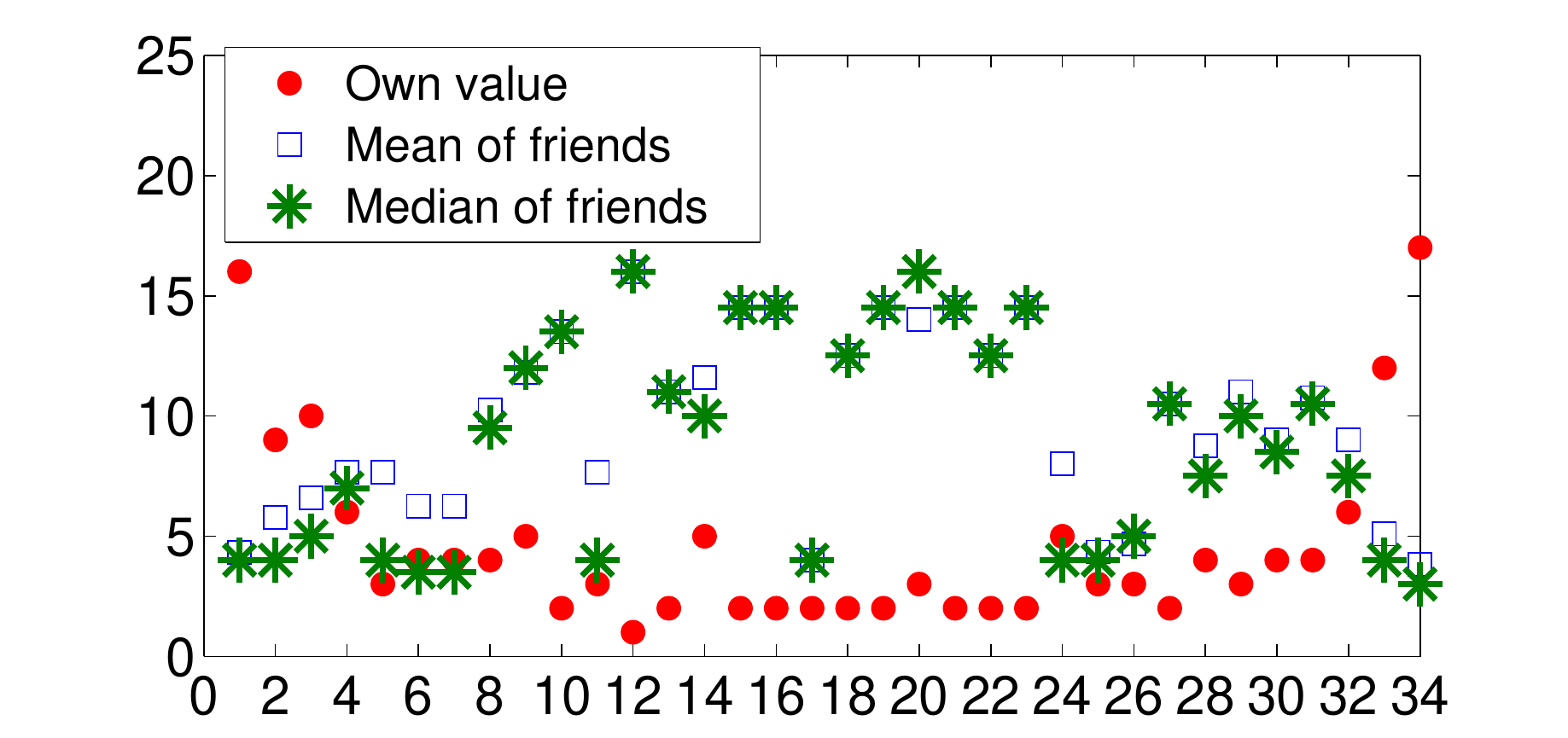}
      \label{fig:karate_paradox}
   }
  \\
   \subfigure[Network paradox for skill. ]{
   \includegraphics[width=0.9\columnwidth]{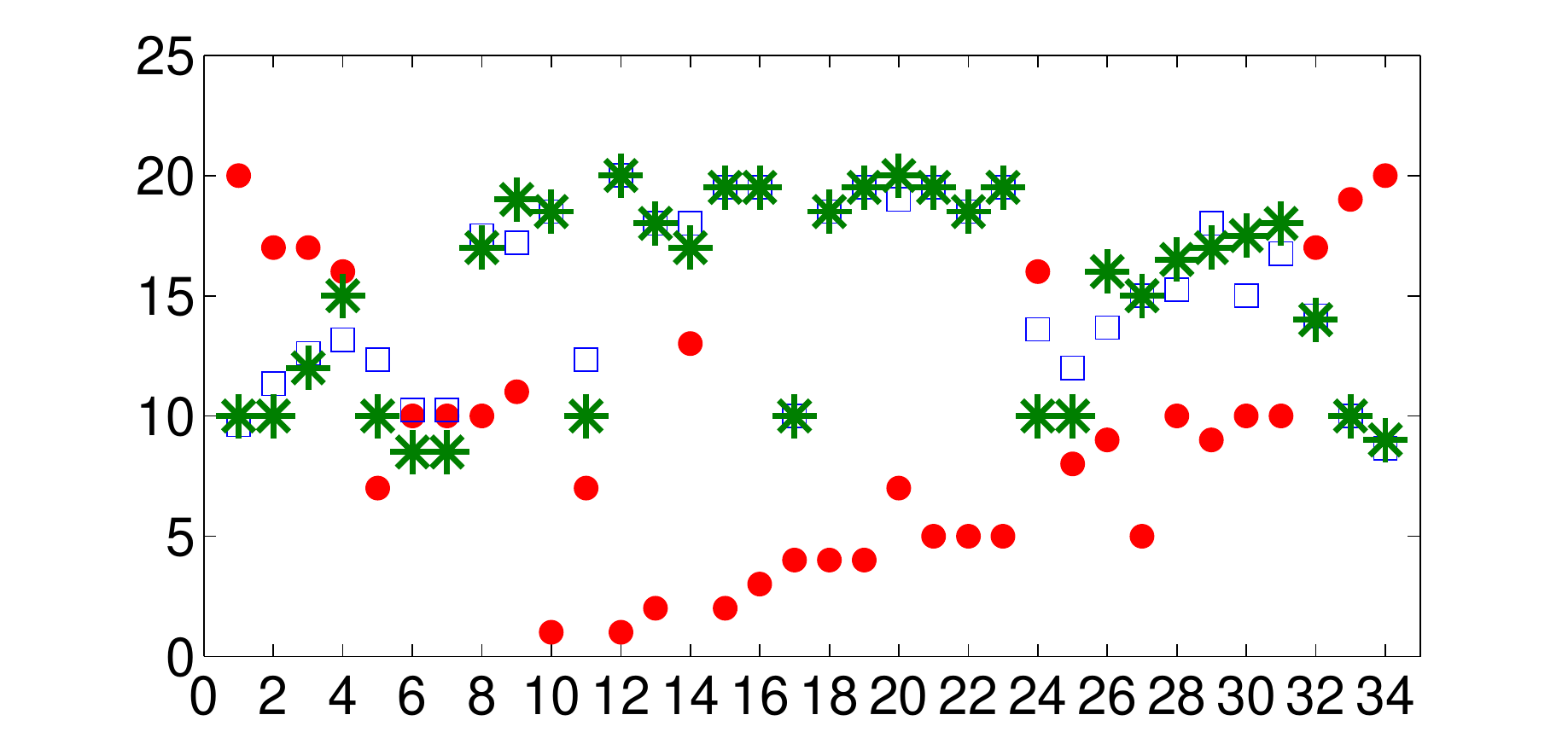}
\label{fig:karate_skill}
   }
  \end{tabular}
  \vspace*{-2mm}
   \caption{Network paradoxes in the Karate Club network. For most individuals, the mean (or median) of the friends' (a) degree and (b) skill level is larger than the individual's own value.}
   \vspace*{-2mm}
\end{figure}

\begin{figure*}[tbh]
\begin{center}
\begin{tabular}{c@{}c@{}c@{}c}
\subfigure[Friends of friends]{%
\label{fig:friend_friend}
\includegraphics[width=0.5\columnwidth]{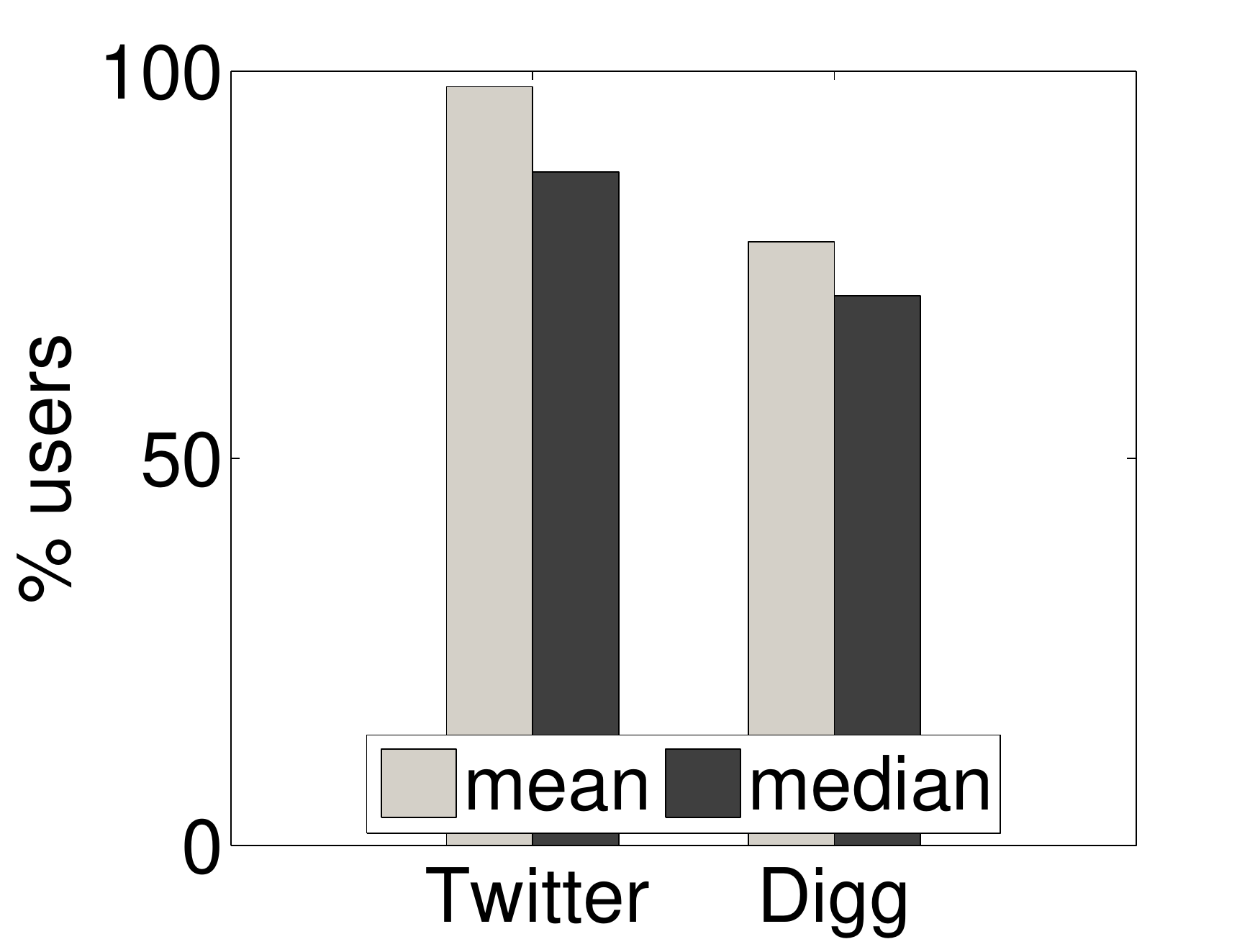}
}
&
\subfigure[Friends of followers]{%
\label{fig:friend_follower}
\includegraphics[width=0.5\columnwidth]{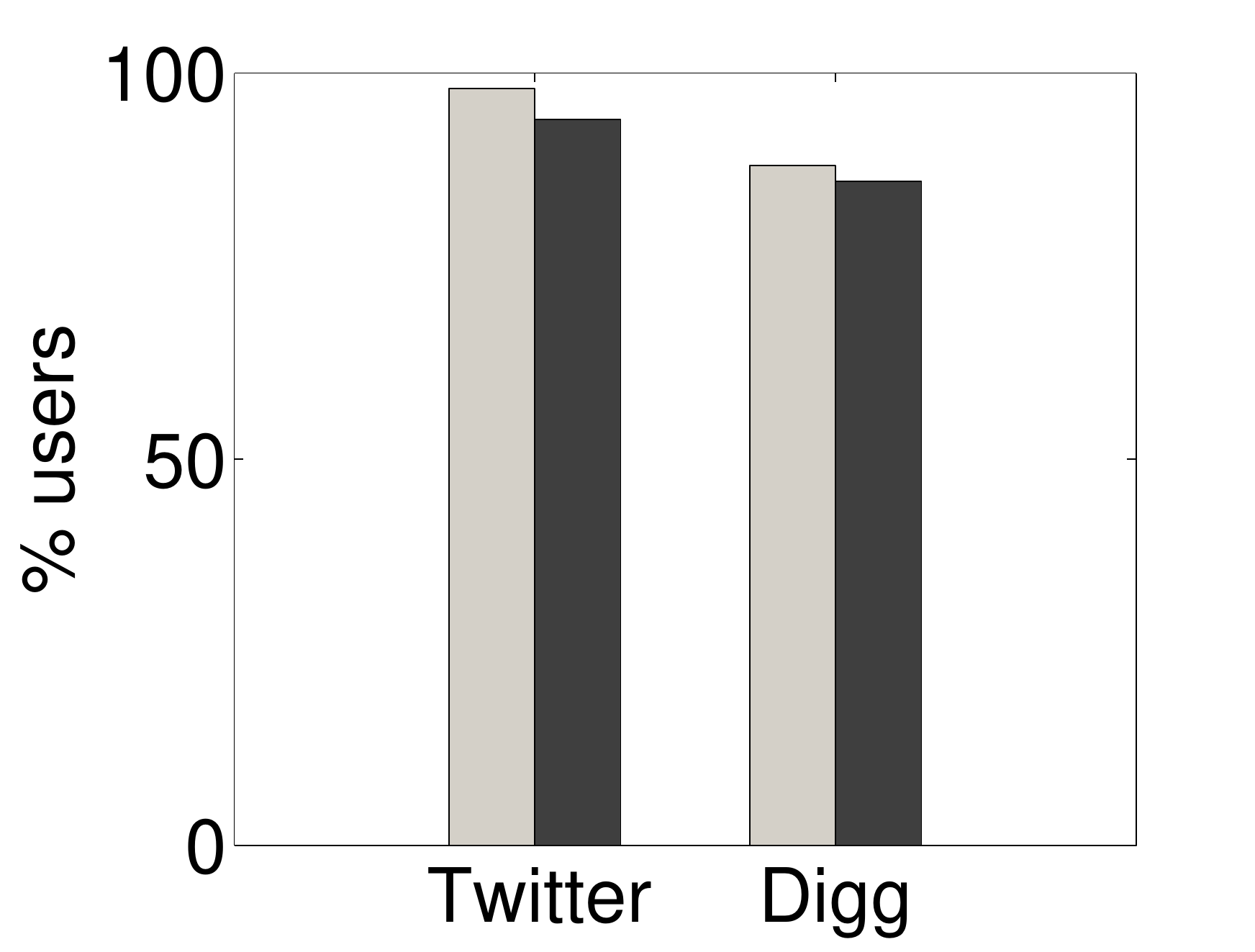}
}
&
\subfigure[Followers of friends]{%
\label{fig:follower_friend}
\includegraphics[width=0.5\columnwidth]{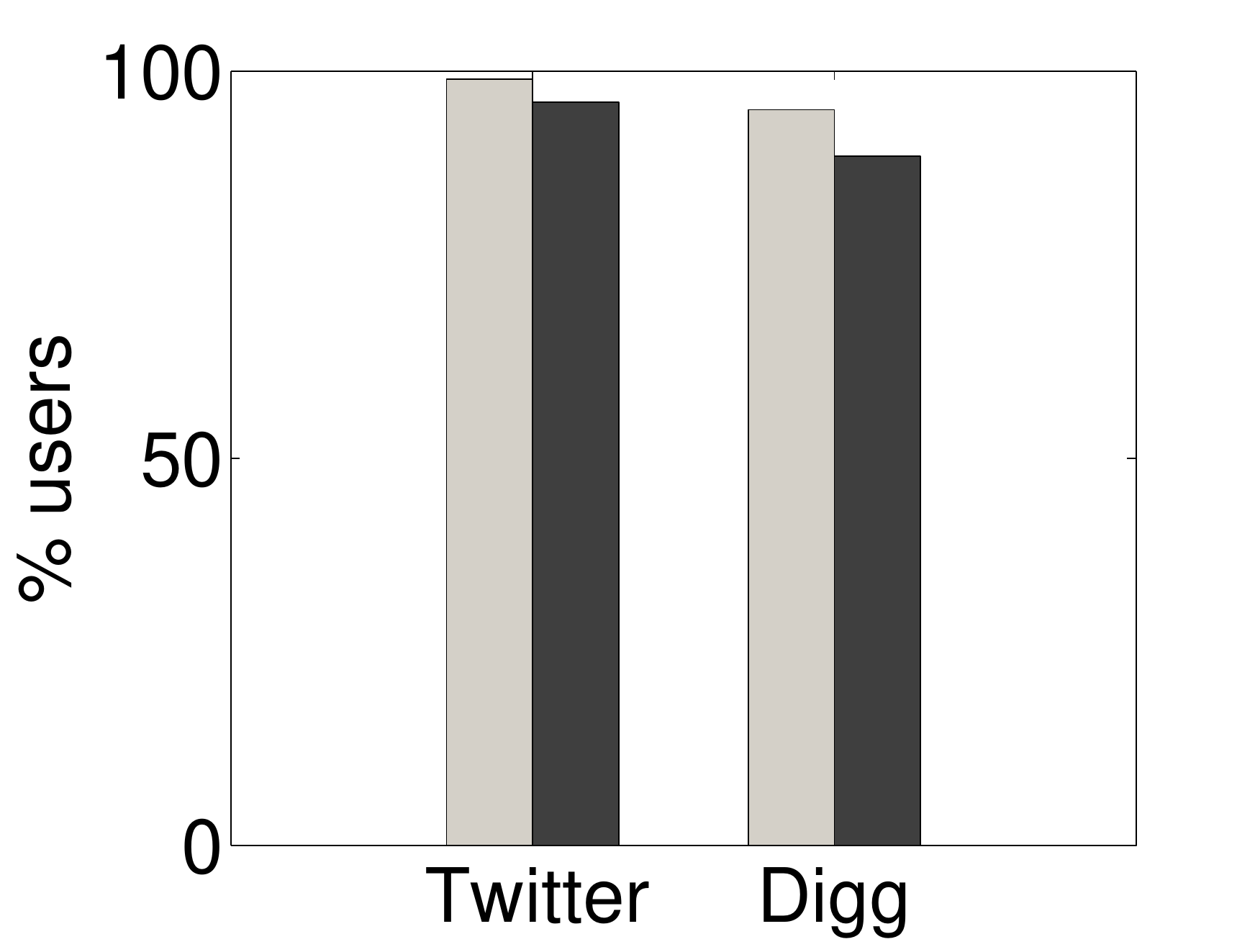}
}
&
\subfigure[Followers of followers]{%
\label{fig:follower_follower}
\includegraphics[width=0.5\columnwidth]{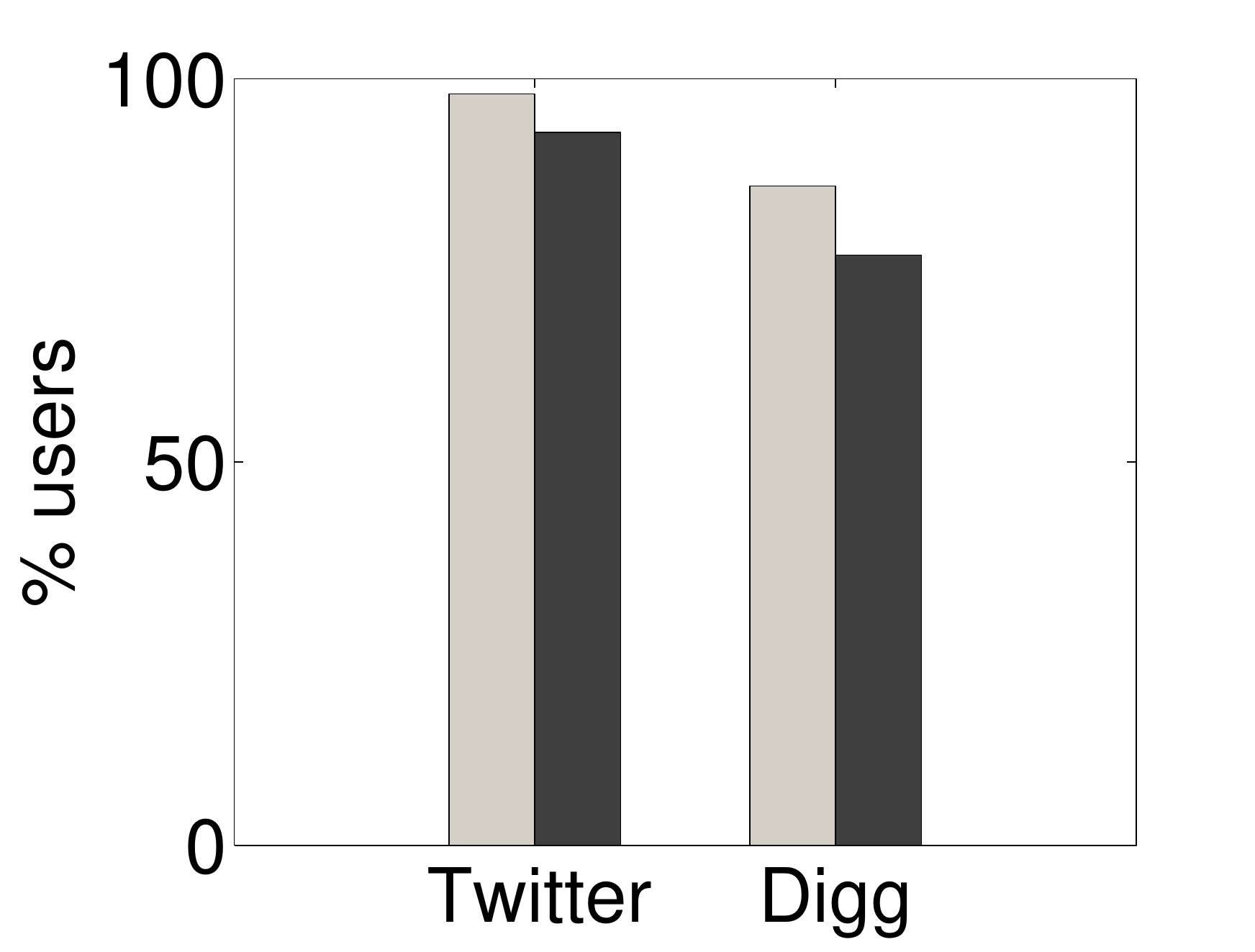}
}
\\
\subfigure[Activity]{%
\label{fig:activity}
\includegraphics[width=0.5\columnwidth]{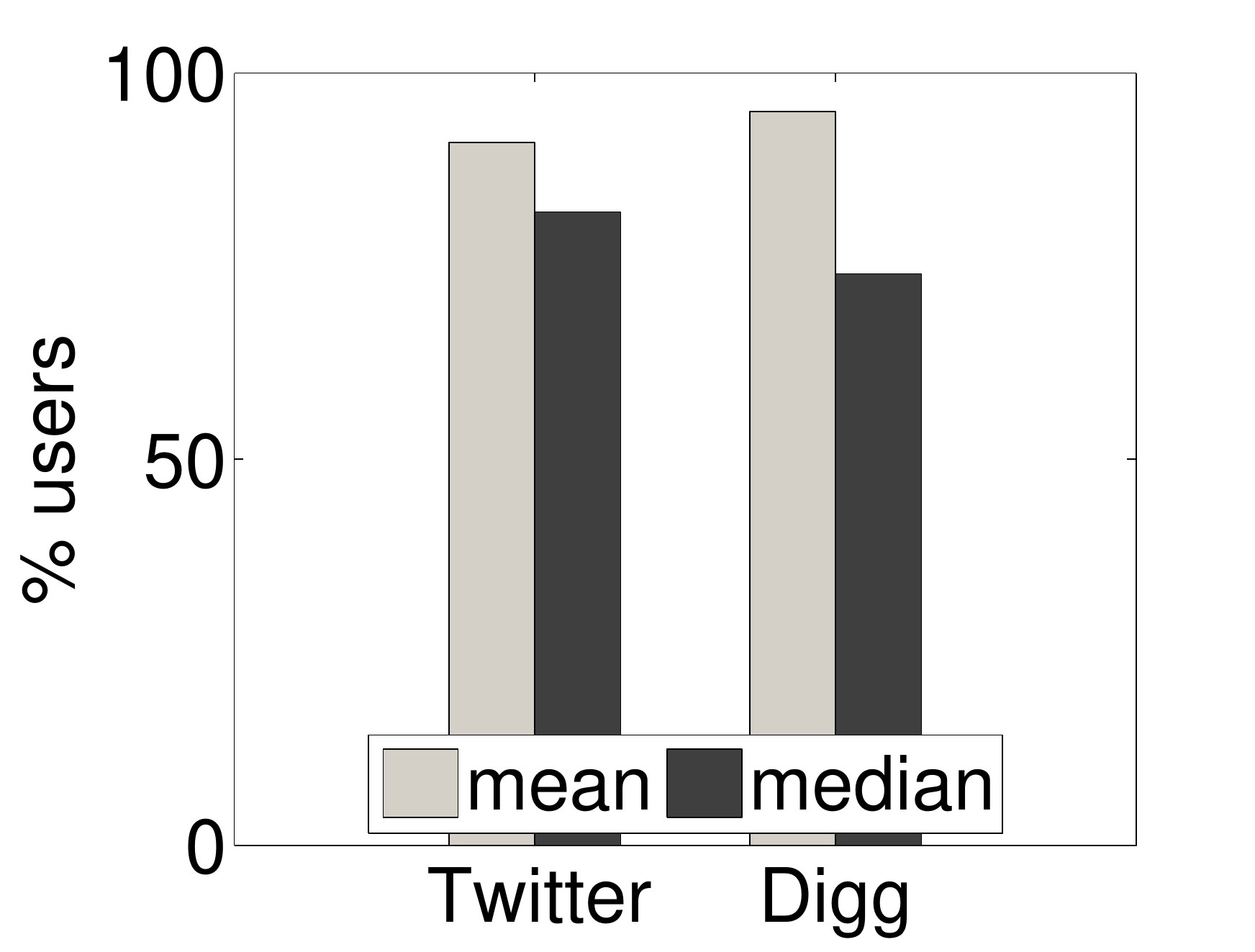}
}
&
\subfigure[Diversity]{%
\label{fig:diversity}
\includegraphics[width=0.5\columnwidth]{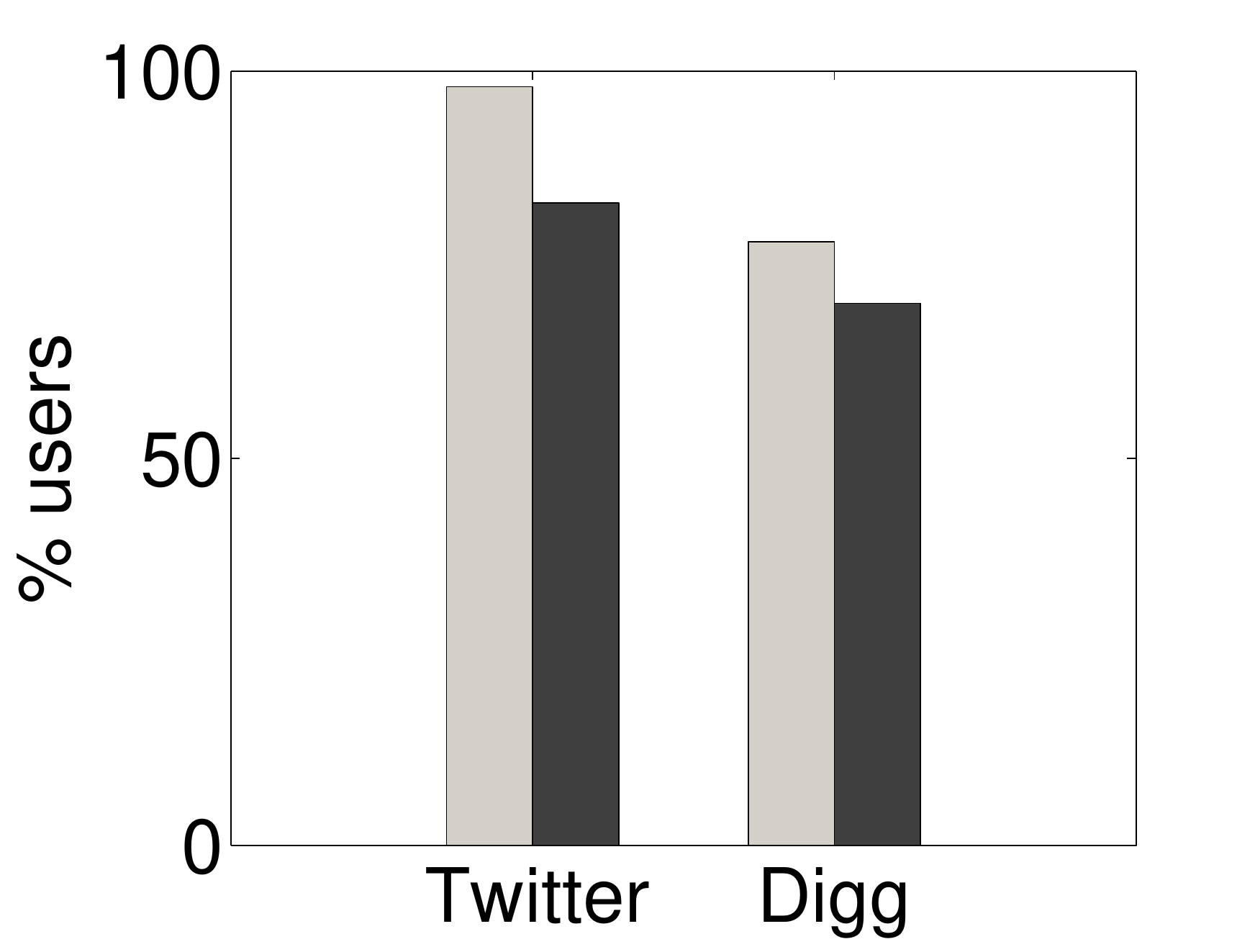}
}
&
\subfigure[Posted URL virality]{%
\label{fig:virality_post}
\includegraphics[width=0.5\columnwidth]{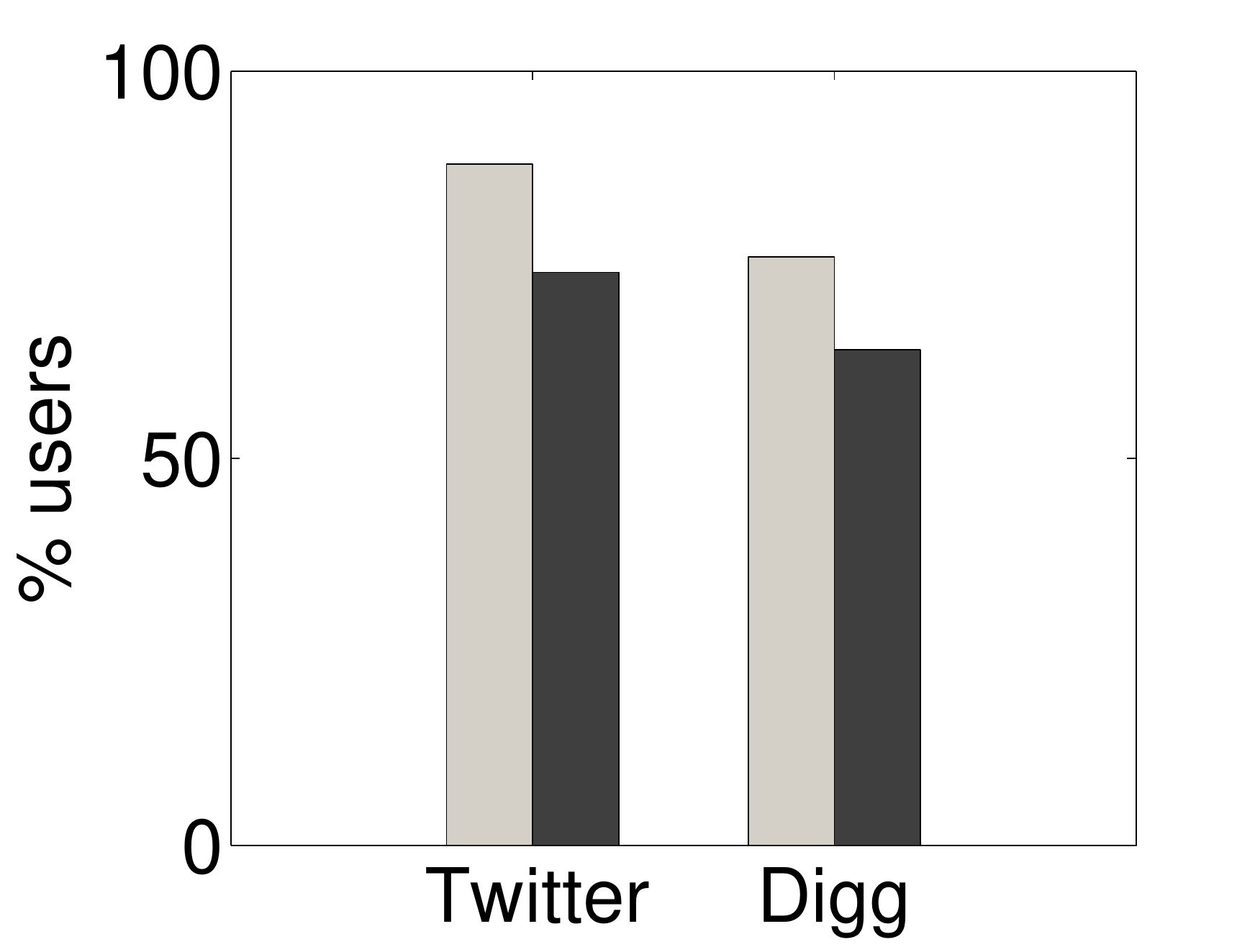}
}
&
\subfigure[Received URL virality]{%
\label{fig:virality_rec}
\includegraphics[width=0.5\columnwidth]{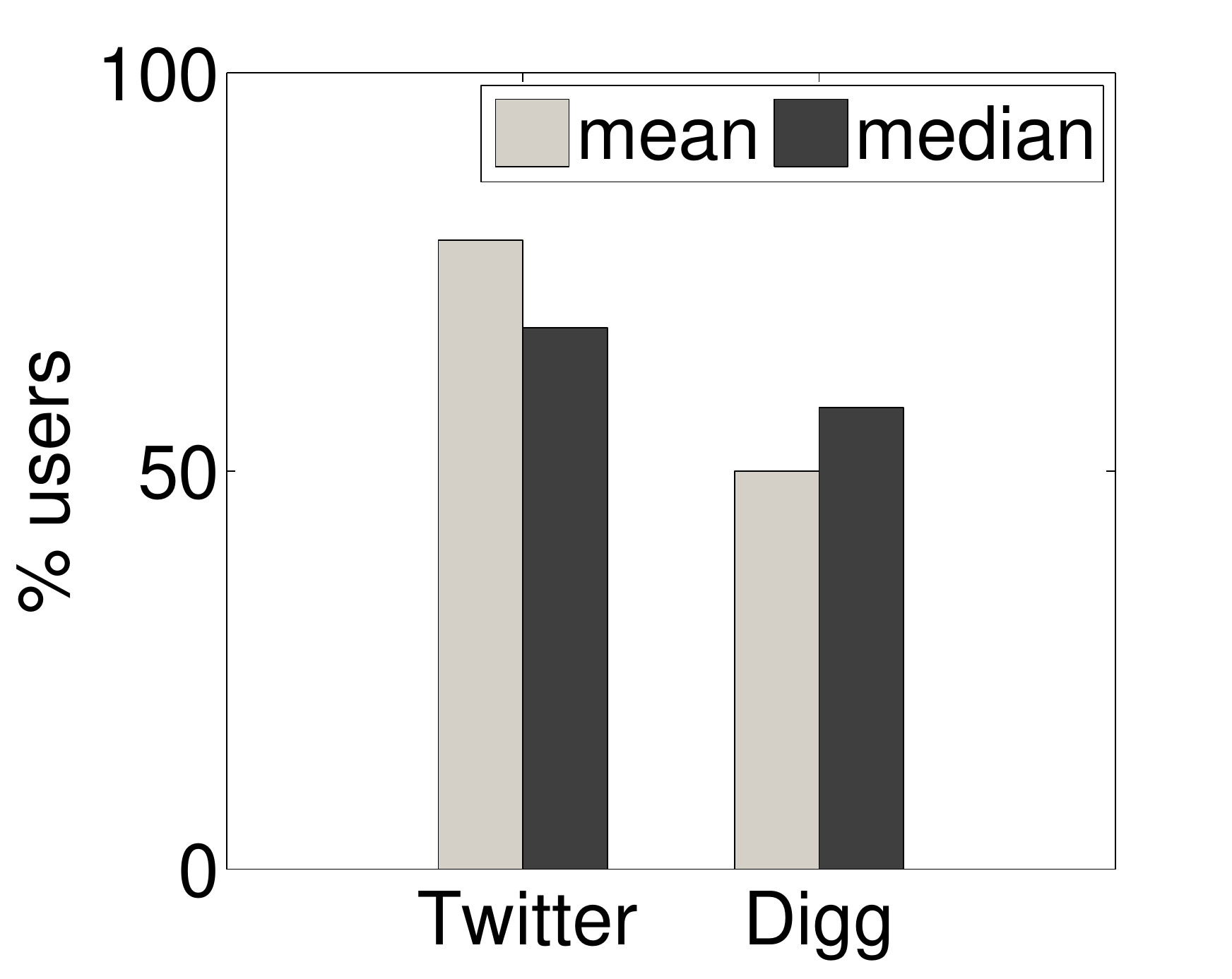}
}
\end{tabular}
\caption{Demonstration of network paradoxes on Twitter and Digg. Each figure shows the percentage of users in the paradox regime, i.e., when the friends's mean (median) attribute value is larger than the user's own attribute value.}
\vspace*{-2mm}
\label{fig:paradoxes}
\end{center}
\vspace*{-2mm}
\end{figure*}

When measuring the paradox for some attribute $x$, we consider a node to be in the paradox regime if the average of the neighbors' values of $x$ is larger than the node's own value. We state that a paradox exists for that attribute if most of the nodes are in the paradox regime.

We demonstrate the network paradoxes in the Karate Club, a small benchmark social network~\cite{zachary1977information}. The network, shown in Figure~\ref{fig:karate_network}, includes 34 individuals and the friendship links between them. There are two nodes, 1 and 34, with a high degree, whereas most of the other nodes have just a few connections.
We calculate the degree of each node along with mean and median degree of the friends,  to demonstrate the friendship paradox. We observe that for 29 out 34 nodes (85\%), the friends have a higher mean degree, and 26 of the nodes (76\%) friends also have a higher median degree (Figure~\ref{fig:karate_paradox}).  Therefore, both weak and strong friendship paradoxes hold for most nodes in this network.

We also consider network paradoxes for node attributes. In the Karate Club data, there are no specific attributes assigned to the individuals, but to demonstrate the network paradoxes, we assign a hypothetical skill level to each individual. We assume that the skill level is correlated with individual's degree. We create skill levels by generating random numbers from a uniform distribution with range of 1-20, and assign the largest number to the node with the largest degree and so on. Similar to the friendship paradox, an attribute paradox exists in the network: for most individuals, their friends have a higher mean (and median) skill level than the individual herself (Figure~\ref{fig:karate_skill}).

Several studies have confirmed that friendship and other network paradoxes exist in a variety of social networks, including Twitter~\cite{garcia2012using,Hodas13icwsm} and scientific collaboration networks~\cite{Eom14}. These paradoxes were measured for the \emph{mean} value of the friends' attributes.
Surprisingly, these paradoxes also exist for the \emph{median}.

First, we consider the friendship paradox. Since both Twitter and Digg are directed networks, the friendship paradox is manifested in four different ways~\cite{garcia2012using,Hodas13icwsm}: (\emph{i}) On average, your friends have more friends than you do. (\emph{ii}) On average your friends have more followers than you do. (\emph{iii}) On average, your followers have more friends than you do. Finally, (\emph{iv}) on average, your followers have more followers than you do.
Figures~\ref{fig:friend_friend}--\ref{fig:follower_follower} show the percentage of users in the paradox regime in each case. The paradox holds for almost all users when considering both the mean and the median, although the fraction of users in the paradox regime is slightly lower for the median. Thus, on both Digg and Twitter, a stronger statement of the friendship paradox holds.
\begin{quote}
     \textbf{\textit{Strong friendship paradox:}} \emph{The majority of your friends (and followers) have more friends (and followers) than you do.}
\end{quote}

Next, we consider network paradoxes for other user attributes besides degree. The activity paradox~\cite{Hodas13icwsm} compares the number of tweets posted by a user on Twitter (or votes  on Digg) with the average number of tweets (votes) made by her friends. Again, the paradox holds for a large fraction of users, both for the mean and the median (Figure~\ref{fig:activity}). %This means that not only are your friends more active on average, but also that most of them are more active than you.
We also observe network paradoxes for the virality of posted (Fig.~\ref{fig:virality_post}) and received (Fig.~\ref{fig:virality_rec}) content and the diversity (Fig.~\ref{fig:diversity}) of the received content on Twitter and Digg. The paradoxes exist regardless of whether the mean or the median is used to measure friends' values. This leads us to restate the paradoxes in their stronger form.
\begin{quote}
\begin{description}
    \item[\textit{Strong activity paradox:}] \emph{The majority of your friends are more active than you are.}
    \item[\textit{Strong diversity paradox:}] \emph{The majority of your friends receive more diverse content than you do.}
    \item[\textit{Strong virality paradox:}] \emph{The majority of your friends send and receive more viral content than you do.}
     \end{description}
\end{quote}

Remarkably, the paradoxes exist for attributes that are beyond user's direct control. Specifically, both diversity and virality of received content depend on the decisions friends make about the content they post. Somehow users position themselves in the network in a way that leads to a paradox: for most of the users in the network a randomly picked friend of the user is highly likely to be better connected, more active, and receive better content. This suggests that if your goal is exposure to interesting and novel content, a promising strategy for identifying new people to follow on Twitter would be to pick a random friend of a user you meant to follow, rather than following that user.
%Higher virality of content posted by friends can be explained by the friendship paradox: since friends also have more followers, their posts are more likely to be retweeted, giving them higher virality.

As explained earlier in this paper, it is not surprising to observe a paradox in the mean for an attribute with a heavy-tailed distribution. Any attribute with such a distribution will manifest the paradox, even in the absence of underlying behavioral factors. As an example, consider the fraction of posted URLs that point to YouTube videos. We do not expect this attribute to be ``paradoxical'', since it is unlikely that users selectively link to others who post a higher fraction of YouTube videos. However, 98\% of users appear to be in the paradox regime when the mean is used. Using the median, on the other hand, puts only 43\% of users in the paradox regime; hence, there is no true paradox -- no surprising behavior. The apparent ``paradox'' only exists \remove{because a few friends with extreme attribute values skew the mean} because the underlying distribution has a mean greater than the median.
\remove{Thus, the median, on the other hand, is more robust to these aberrations and should be used as an indicator of paradoxes.}

Network paradoxes are not a manifestation of Simpson's paradox. Simpson's paradox refers to a phenomenon when a specific trend is observed within different sub-groups of data, but the trend does not appear if these groups are combined. For example, the mean within  sub-groups could be above a threshold, but the aggregate mean over all groups is below the threshold. Simpson's paradox arises from \remove{the grouping of data} mixing of heterogeneous populations, but in our case we are not  grouping users based on any attribute. We simply check each individual to see whether she is in a paradox regime and report the fraction of individuals in the paradox regime.

\section{Behavioral Origins of Paradoxes}
\label{sec:shuffletest}

%As argued in this paper, network paradoxes for the median do not arise purely due to statistical properties of the distribution of user-attribute values, but they must be due to some behavioral mechanisms. 
We test two potential sources of the behavioral mechanisms. The first source is the correlation between a user's degree and her own attributes. We call this ``within-node correlation''. We use Pearson's Correlation Coefficient to measure the within-node correlation between its number of friends and its attribute, as defined in \cite{Eom14}. We use the number of friends as the degree and not the number of followers, because only the former characteristic is under user control.
The second potential source of the paradoxes is the correlation between an attribute of the node and the attributes of its neighbors. This correlation is at the link level, and we call it ``between-node correlation''. We use assortativity to measure this correlation. Table~\ref{table:properties} reports the empirical values (\emph{Emp.}) of assortativity and correlation for a variety of attributes in the Twitter and Digg networks. Note that the follower graphs of Twitter and Digg have a slight degree disassortativity, as found by a previous study of Twitter~\cite{Kwak10www}. Other attributes, on the other hand, are somewhat assortative. Within-node correlations are higher, as observed also in co-authorship networks~\cite{Eom14}.

\begin{table}[th!]
\begin {center}
\scalebox{0.8} {
\begin {tabular} {|l@{}|c|c|c||c|c|c|}
\hline
 \multicolumn{7}{|c|}{\textsc{Assortativity of the attribute}}\\ \hline
\textbf{Attribute} & \multicolumn{3}{|c||}{\textbf{Twitter}} & \multicolumn{3}{|c|}{\textbf{Digg}} \\ \hline
& \emph{Emp.} & \emph{Contr.} & \emph{Shuffle} & \emph{Emp.} & \emph{Contr.}  & \emph{Shuffle} \\
\hline
num. friends  & 0.015 & --- & 0.000 & -0.040 & --- & 0.001  \\
num. followers & -0.047&--- & 0.000& -0.157 & --- & 0.001  \\
activity & 0.037& 0.016& 0.000& 0.152 & 0.005 & 0.000  \\
diversity  & 0.055& 0.012 & 0.000& -0.041& 0.022 & -0.001  \\
posted virality & 0.030 & 0.000 & 0.000& 0.061 & 0.000 &  0.000 \\
received virality & 0.191 & 0.001& 0.000& 0.105 & 0.010 & 0.000  \\
\hline
%\end{tabular}}
%\end{center}
%\vspace*{-2mm}
%\caption{Assortativity of attributes of connected users calculated in the empirical data, partially shuffled shuffled, and completely shuffled networks.}
%\end{table}
%\begin{table}[th]
%\begin {center}
%\scalebox{0.8} {
%\begin {tabular} {|l@{}| c | c | c || c | c | c |}
%\hline
% & \multicolumn{3}{|c||}{\bf Twitter} & \multicolumn{3}{|c|}{\bf Digg} \\
\hline
\multicolumn{7}{|c|}{\textsc{Correlation of node's attribute with num. friends}}\\ \hline
\textbf{Attribute} & \multicolumn{3}{|c||}{\textbf{Twitter}} & \multicolumn{3}{|c|}{\textbf{Digg}} \\ \hline
 & \emph{Emp.} & \emph{Contr.} & \emph{Shuffle} & \emph{Emp.} & \emph{Contr.}  & \emph{Shuffle} \\
\hline
activity & 0.191& 0.138& -0.001& 0.097& 0.108 & -0.002 \\
diversity & 0.895& 0.867& 0.001& 0.999& 0.690&  0.005\\
posted virality & -0.001& -0.001& 0.000& -0.019& 0.040& 0.003\\
received virality & 0.000& 0.000& 0.000& 0.287& 0.281& 0.001\\
\hline
\end{tabular}
}
\end{center}
\vspace*{-2mm}
\caption{Network properties. (\emph{Top}) Assortativity of attributes of connected users and (\emph{Bottom}) within-node correlations of the attribute with degree in the empirical data (\emph{Emp.}) and in the shuffled networks after a controlled (\emph{Contr.}) and full shuffle (\emph{Shuffle}) of attributes. }\label{table:properties}
\end{table}

%In this section we investigate alternate explanations for paradoxes, specifically, whether it is caused by within-node correlations or assortativity or both.
We use the shuffle test to probe the behavioral explanation of the paradoxes. The shuffle test  randomizes node attribute values, destroying  the within-node and/or between-node correlations. We then measure network paradox for the attribute  in the shuffled network. If the paradox disappears, because most of the users are no longer in the paradox regime, we conclude that the correlation is the root cause of the paradox. First, we start by destroying both correlations and observe that the strong form of the paradoxes disappear, so the paradoxes are caused by these correlations. Then, we do a controlled shuffle to differentiate between within-node and between-node correlations.

\subsection{Shuffle Test}

\begin{figure}[tb]
\begin{center}
\begin{tabular}{c@{}c}
\subfigure[Friends of friends]{%
\label{fig:friend_friend_shuff}
\includegraphics[width=0.5\columnwidth]{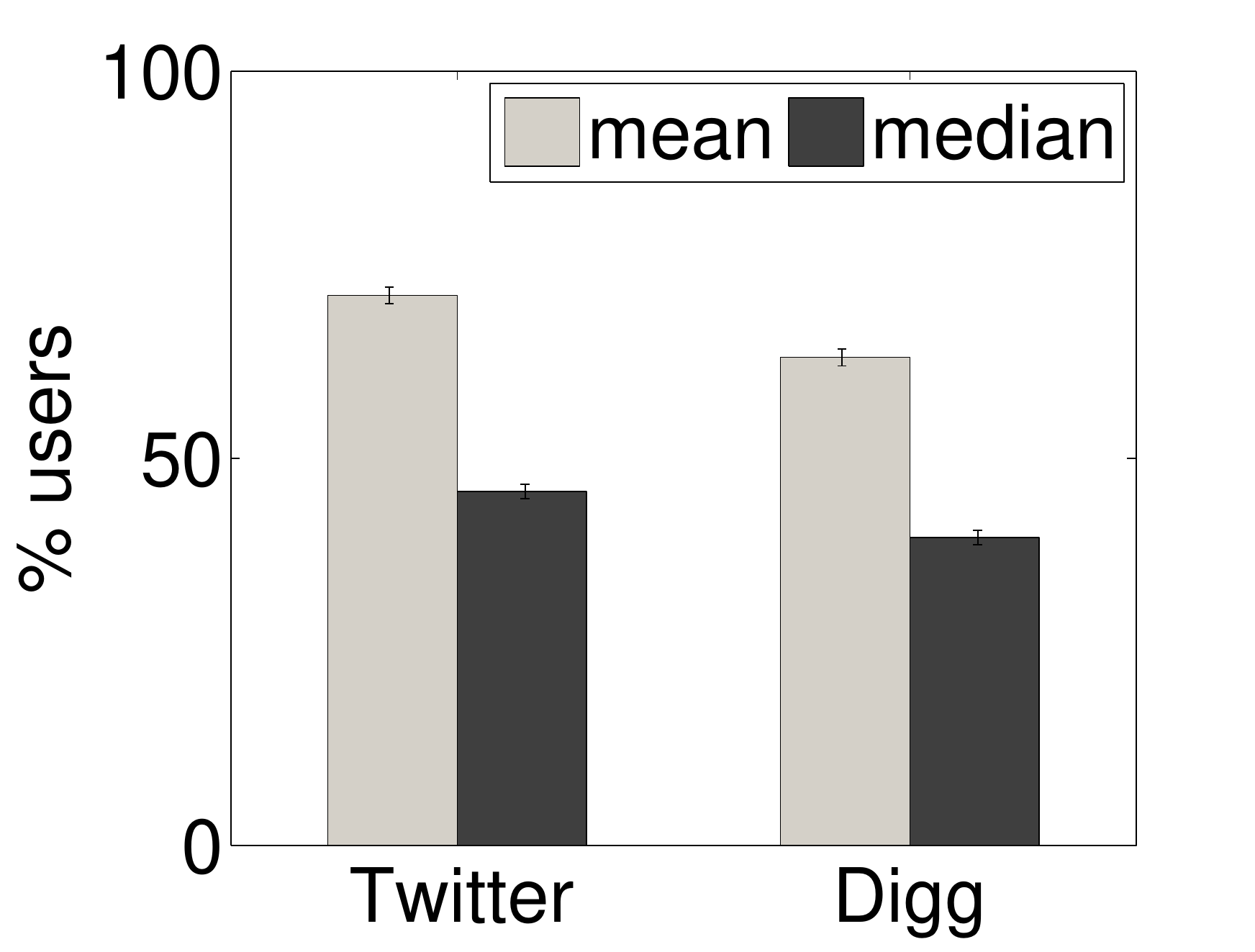}
}
&
\subfigure[Friends of followers]{%
\label{fig:friend_follower_shuff}
\includegraphics[width=0.5\columnwidth]{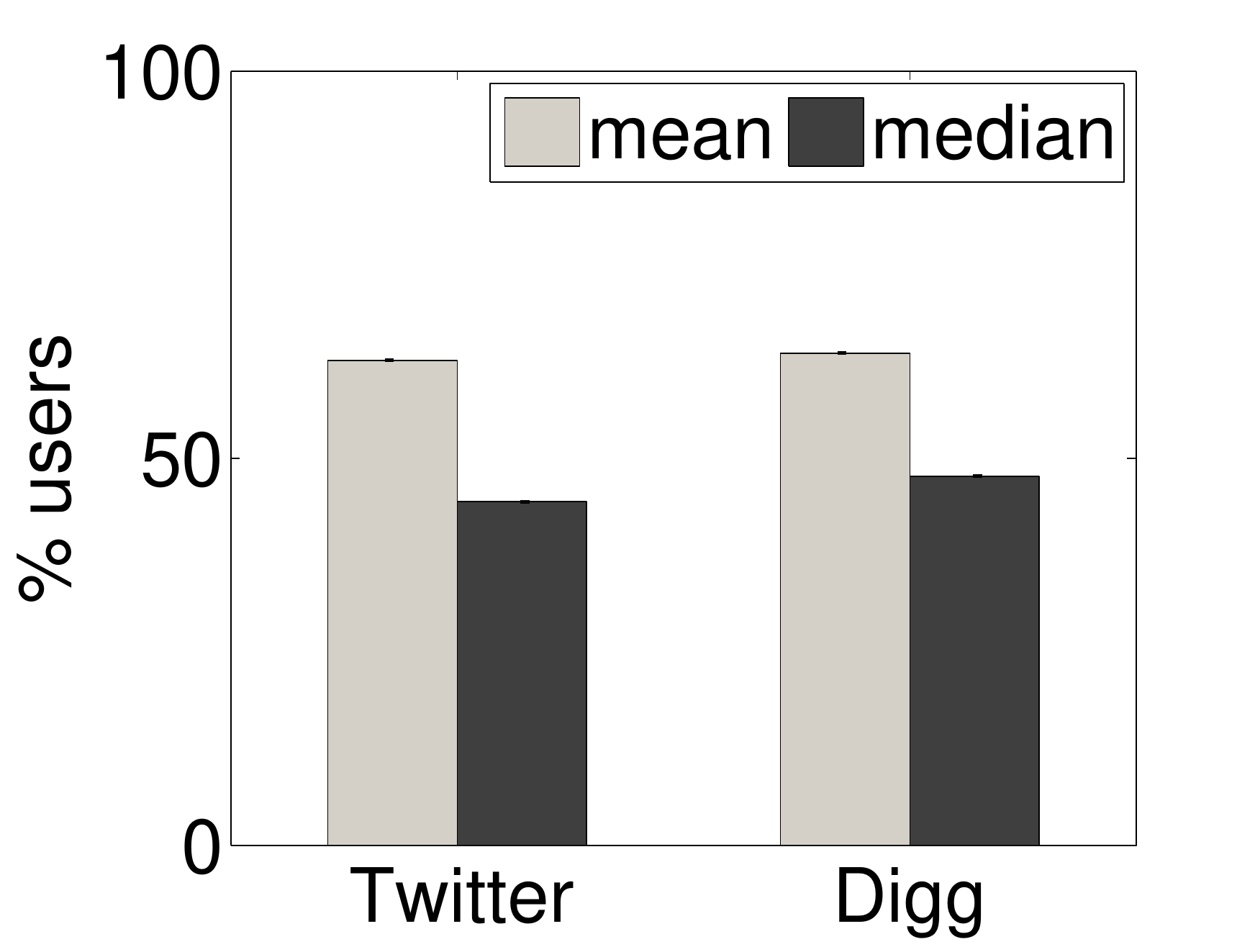}
}
\\
\subfigure[Followers of friends]{%
\label{fig:follower_friend_shuff}
\includegraphics[width=0.5\columnwidth]{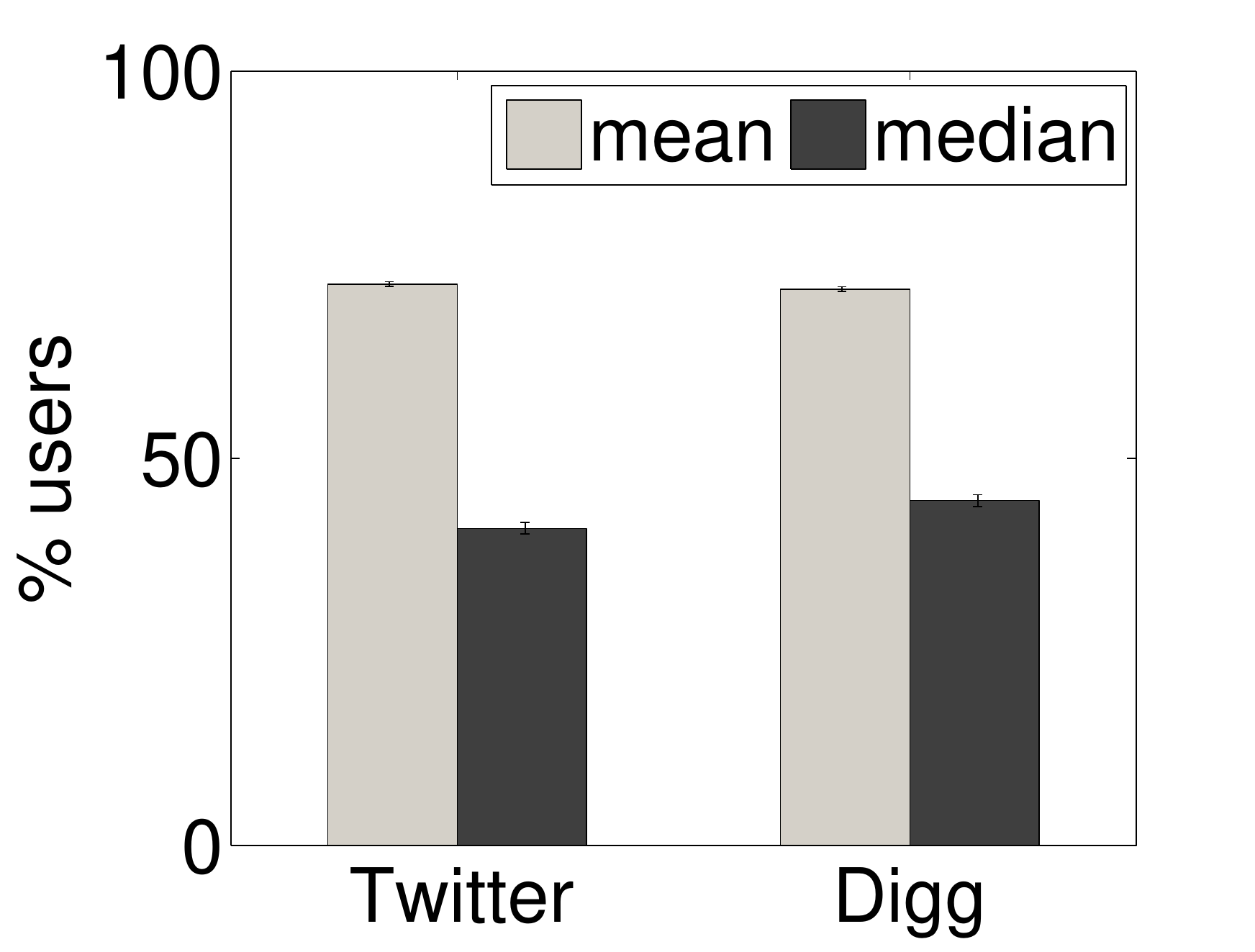}
}
&
\subfigure[Followers of followers]{%
\label{fig:follower_follower_shuff}
\includegraphics[width=0.5\columnwidth]{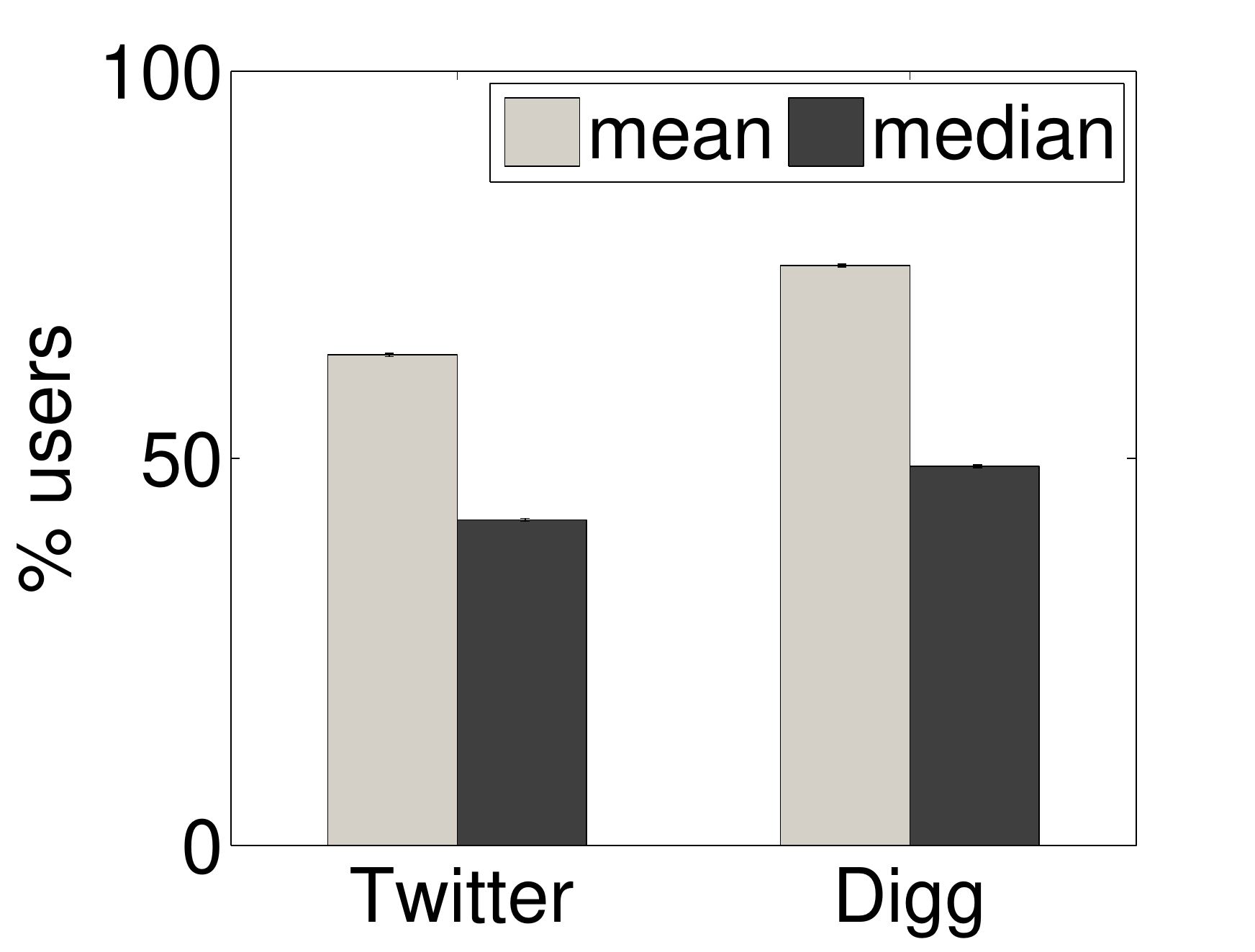}
}
\end{tabular}
\vspace*{-2mm}
\caption{Percentage of users in paradox regime on Twitter and Digg after shuffling the number of  friends (top row) and the number of followers (bottom row). Error bars show the 0.95 confidence interval.}
\label{fig:friend_shuffle}
\vspace*{-2mm}
\end{center}
\end{figure}

%\paragraph{Degree}
We start by examining the number of friends attribute. As explained earlier, in directed networks there are four variants of the friendship paradox, which compare the number of friends or followers a user has with the average number of friends or followers of her (\emph{i}) friends and (\emph{ii}) followers.
%To check whether each paradox arises due to correlation of the number of friends attribute of connected users,\footnote{Users can be connected through a friend or follower relationship.}
We shuffle the network to destroy the correlation between connected nodes as follows.  We keep the links between users as is, preserving network structure, but assign a new number of friends to each user, which is randomly drawn from another network node.
Note that ``number of friends'' is treated as an attribute of a node, unrelated to its degree. While this may be a non-conventional application of the shuffle test, we use it to probe how correlations affect the paradoxes.
Shuffling the number of friends eliminates any correlation between the number of friends of connected users, but does not change its distribution.
Table~\ref{table:properties} (column \emph{Shuffle}) shows that in all cases degree assortativity disappears in the shuffled  network.  Figures~\ref{fig:friend_friend_shuff} and \ref{fig:friend_follower_shuff} show that the friendship paradox still holds in the shuffled network (though weaker) for the mean. However, the paradox no longer holds for the median, since fewer than 50\% of users are in the paradox regime in the shuffled network.
%The presence of the paradoxes for the mean, but not the median confirms the theoretical findings presented in the earlier sections.

%\subsection{Shuffling Followers}
Next, we consider the paradoxes involving the number of followers. We shuffle the number of followers by assigning to each user the number of followers from a randomly drawn user. The two paradoxes still hold for more than 60\% of users for the mean, but only about 50\% of users are in the paradox regime for the median on Twitter and Digg (Figs.~\ref{fig:follower_friend_shuff}--\ref{fig:follower_follower_shuff}).

The empirically observed degree dissassortativity is an outcome of the mechanisms people use to select who to follow in online social networks. Disassortativity appears in a network where below-average users follow above-average users. This seems to be the case on Twitter (and Digg) where large fraction of follow links are from normal users \remove{(with few followers)} to the celebrities (or top users on Digg) with orders of magnitude more followers.
Friendship paradoxes in the online social networks of Digg and to some extent Twitter appear to be related to degree disassortativity.

\remove{
\begin{figure}[tbh]
\begin{center}
\begin{tabular}{cc}
\end{tabular}
\caption{Percentage of users in paradox regime on Twitter and Digg after shuffling the number of  followers. Error bars show the 0.95 confidence interval.}
\label{fig:follower_shuffle}
\end{center}
\end{figure}
}

%\paragraph{Other attributes}
The remaining paradoxes are similar because each compares a user's attribute with the average value of this attribute among her friends. In each shuffle test, we shuffle the values of the attribute among all users. This eliminates both within-node and between-node correlations.
%, because connected users have been assigned random values of the attribute. Also, the shuffle test destroys
%, because each user and her friends have random and uncorrelated values of the attribute.
Table~\ref{table:properties} (column \emph{Shuffle})  shows that none of the correlations exist in the shuffled network.

\begin{figure}[tb]
\begin{center}
\begin{tabular}{@{}c@{}c@{}}
\subfigure[Activity]{%
\label{fig:activity_shuff}
\includegraphics[width=0.5\columnwidth]{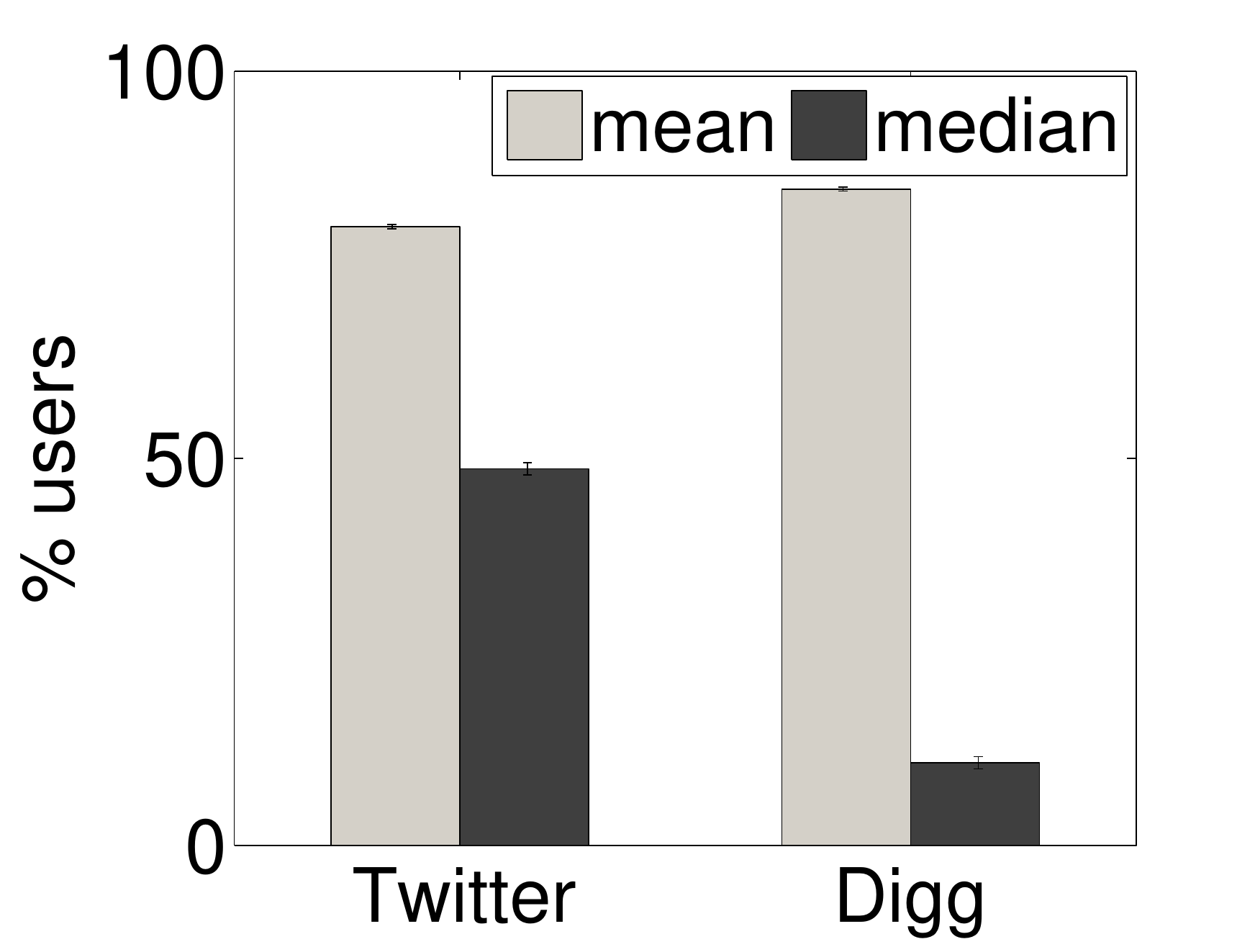}
}
&
\subfigure[Diversity]{%
\label{fig:diversity_shuff}
\includegraphics[width=0.5\columnwidth]{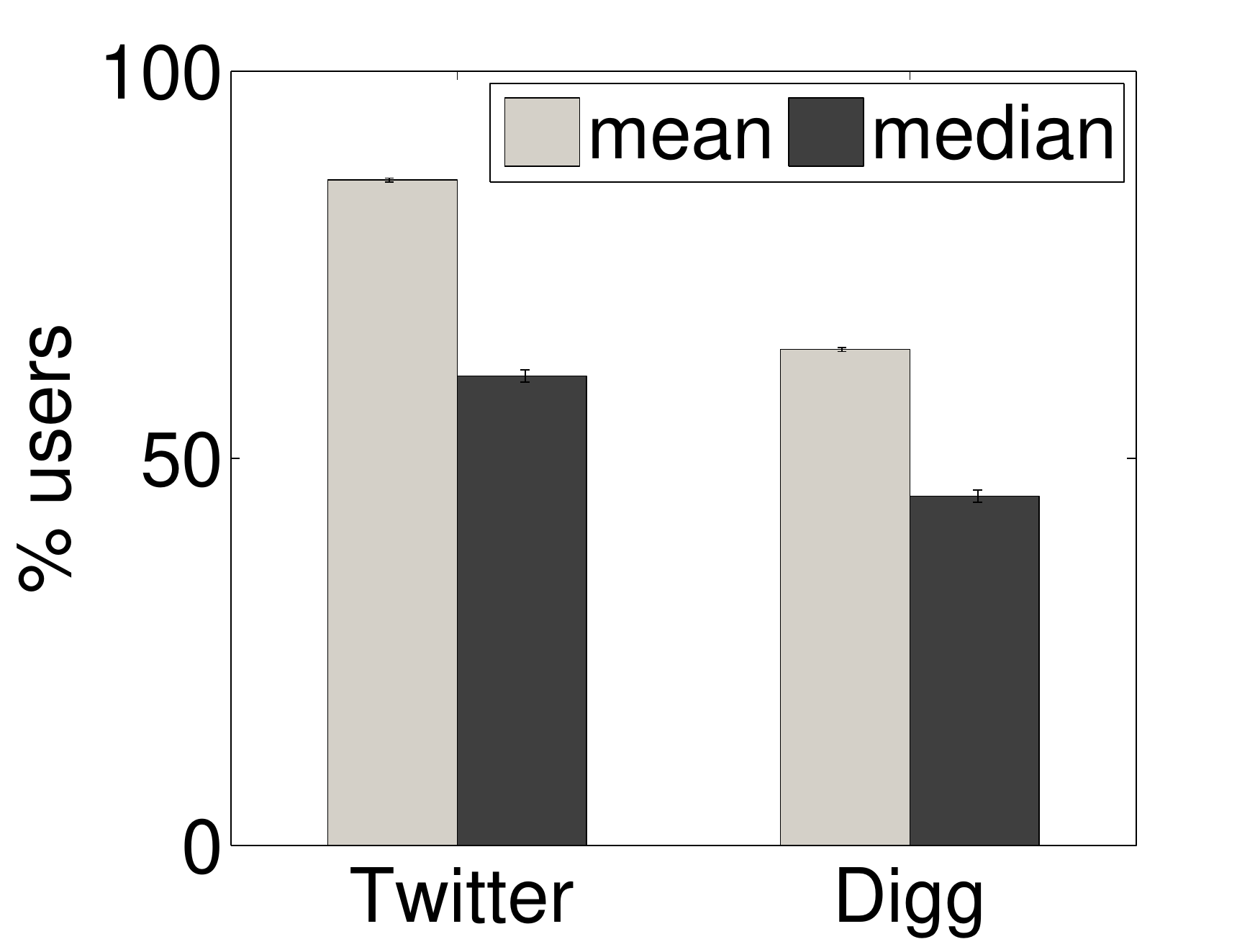}
}
\\
\subfigure[Posted URL virality]{%
\label{fig:virality_post_shuff}
\includegraphics[width=0.5\columnwidth]{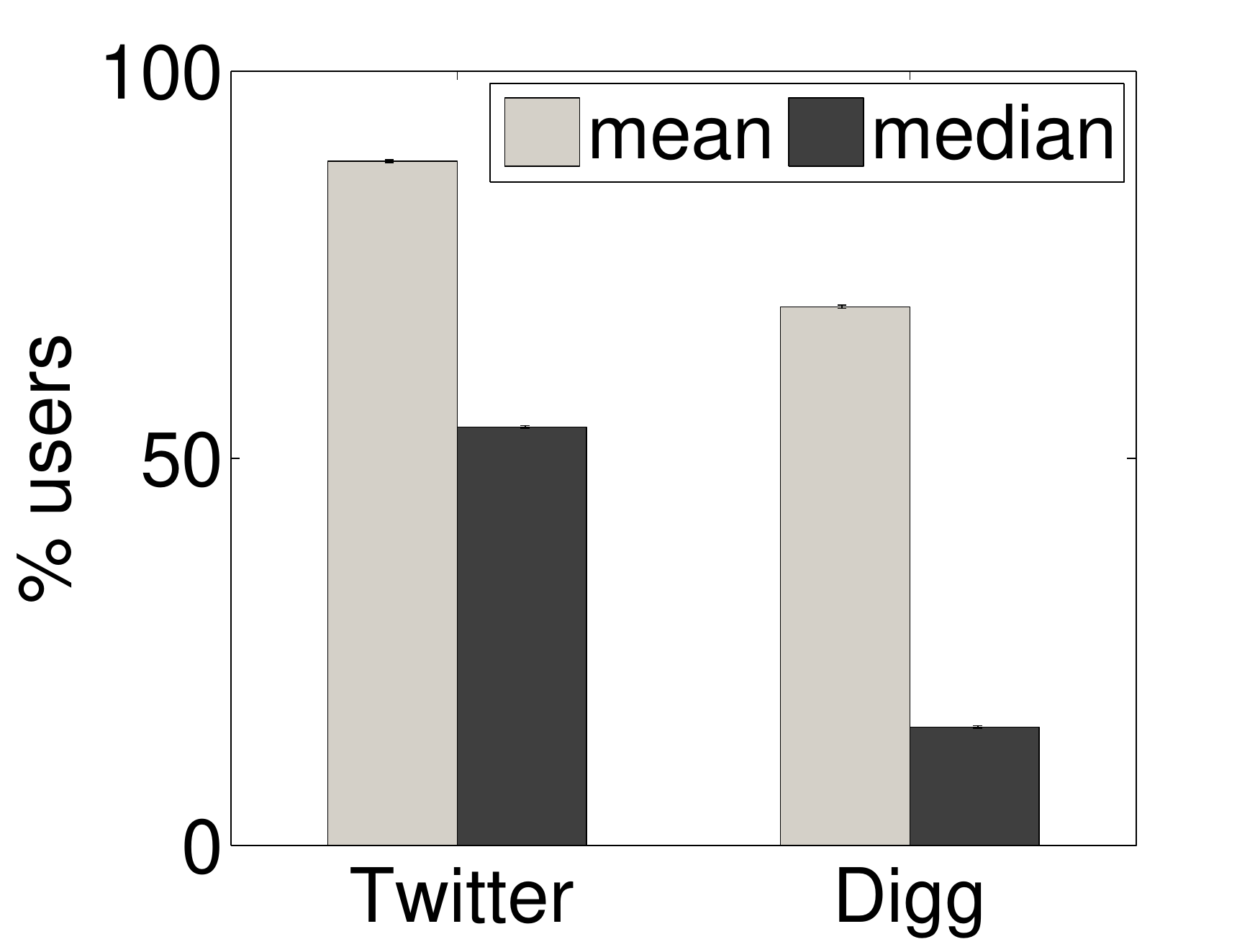}
}
&
\subfigure[Received URL virality]{%
\label{fig:virality_rec_shuff}
\includegraphics[width=0.5\columnwidth]{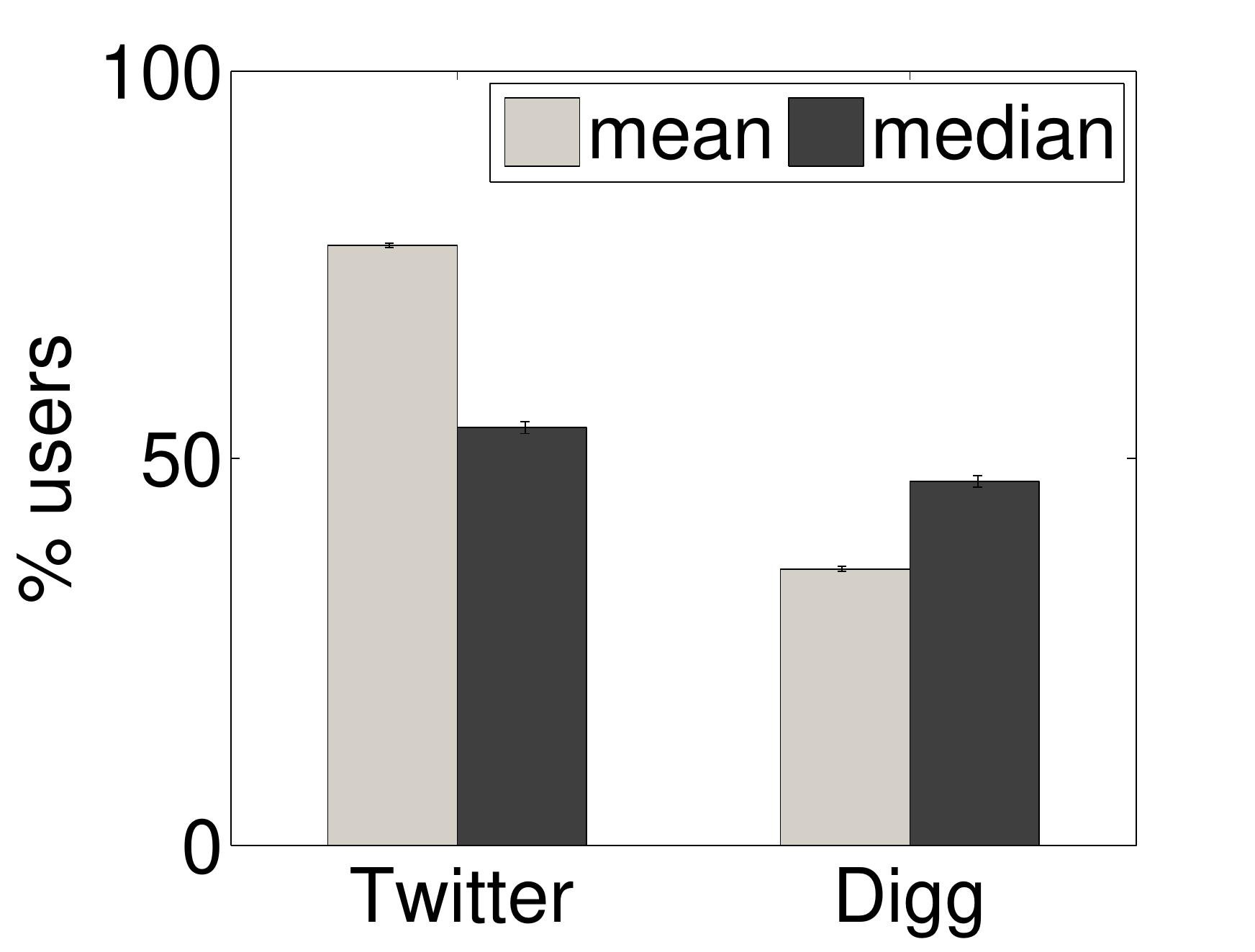}
}
\end{tabular}
\vspace*{-2mm}
\caption{Percentage of users in paradox regime on Twitter and Digg after shuffling user attribute. Error bars show the 0.95 confidence interval.}
\label{fig:attr_shuffle}
\end{center}
\vspace*{-2mm}
\end{figure}

Figure~\ref{fig:attr_shuffle} measures network paradoxes for the four attributes in the shuffled Twitter and Digg networks. In almost all cases, the paradoxes still hold for the mean. The only exception is the received virality paradox on Digg, which does not hold because virality of the stories does not have a heavy-tailed distribution on Digg, as mentioned earlier. When comparing user with friends using the median, the paradoxes mostly disappear. One ambiguous case is content diversity paradox on Twitter, which has 60\% of the users in the paradox regime, a small statistically significant paradox.

We conclude that the empirical observations of the paradox for the median cannot be explained purely by statistical sampling and imply a socio-behavioral dimension. The origin of these strong network paradoxes appears to be in the within- and between-node correlations.

\subsection{Controlled Shuffle Test}
Eom and Jo~\cite{Eom14} argued that within-node correlation between attribute and degree results in the observed paradox in the mean for the attribute. Unfortunately, the shuffle test does not allow us to distinguish whether within-node or between-node correlation (assortativity) is responsible for the paradox. In this section, we disentangle these effects through a controlled shuffle test, which attempts to eliminate the between-node correlation while preserving within-node correlation. We achieve this by grouping together users with the same degree (number of friends) and shuffling the attribute values within each group. Thus, the reassigned attribute is still correlated with degree, because it's from another user with the same degree. We log-bin the data to deal with degree sparseness at high values. Table~\ref{table:properties} shows that the within-node correlation has not changed significantly, but the between-node correlation is reduced in the shuffled network (column \emph{Contr.}).
%, because the re-assigned attribute is randomly chosen from the attributes of a large group of users.

\begin{figure}[tb]
\begin{center}
\begin{tabular}{@{}c@{}c@{}}
\subfigure[Activity]{%
\label{fig:activity_assortativity_shuff}
\includegraphics[width=0.5\columnwidth]{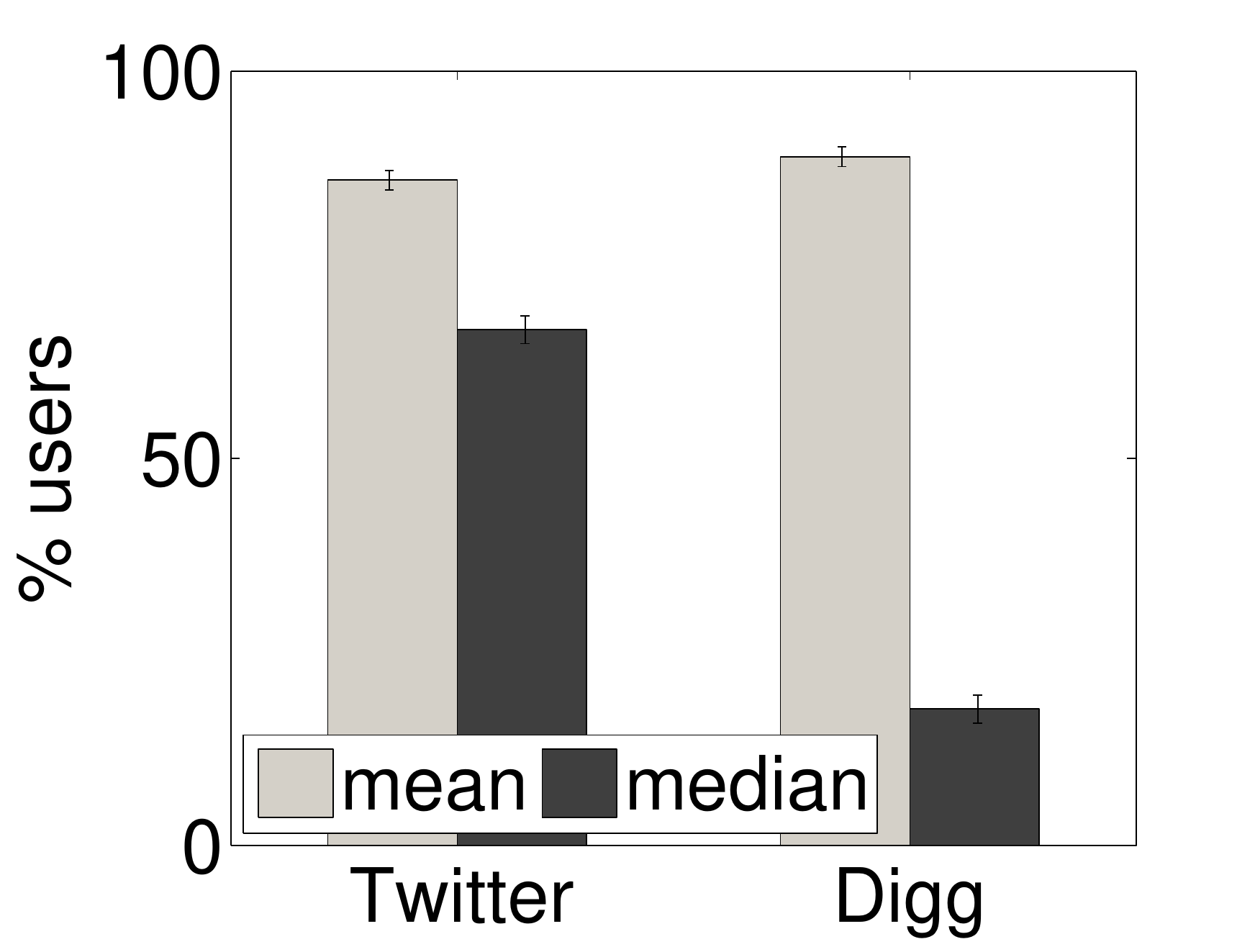}
}
&
\subfigure[Diversity]{%
\label{fig:diversity_assortativity_shuff}
\includegraphics[width=0.5\columnwidth]{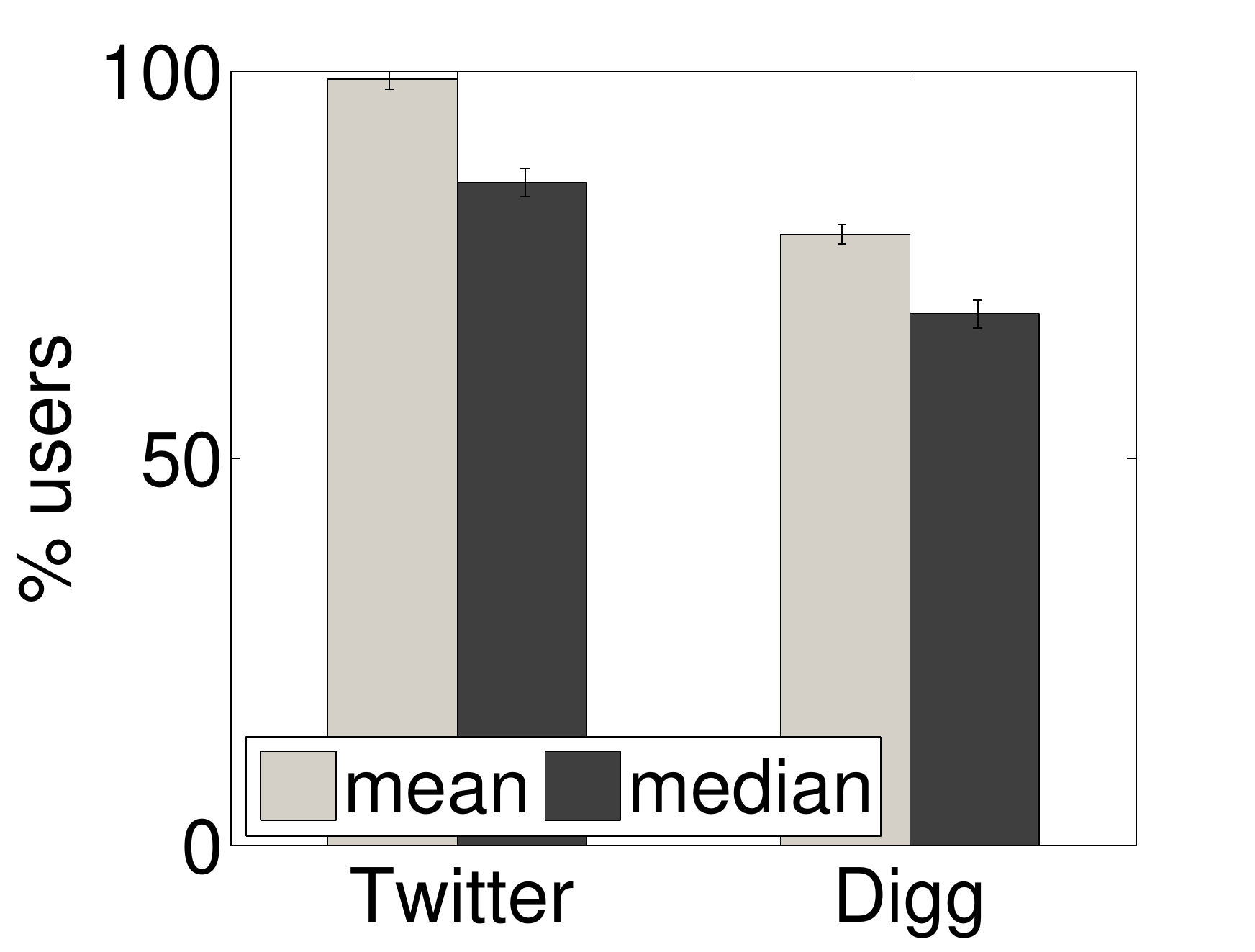}
}\\
\subfigure[Posted URL virality]{%
\label{fig:virality_post_assortativity_shuff}
\includegraphics[width=0.5\columnwidth]{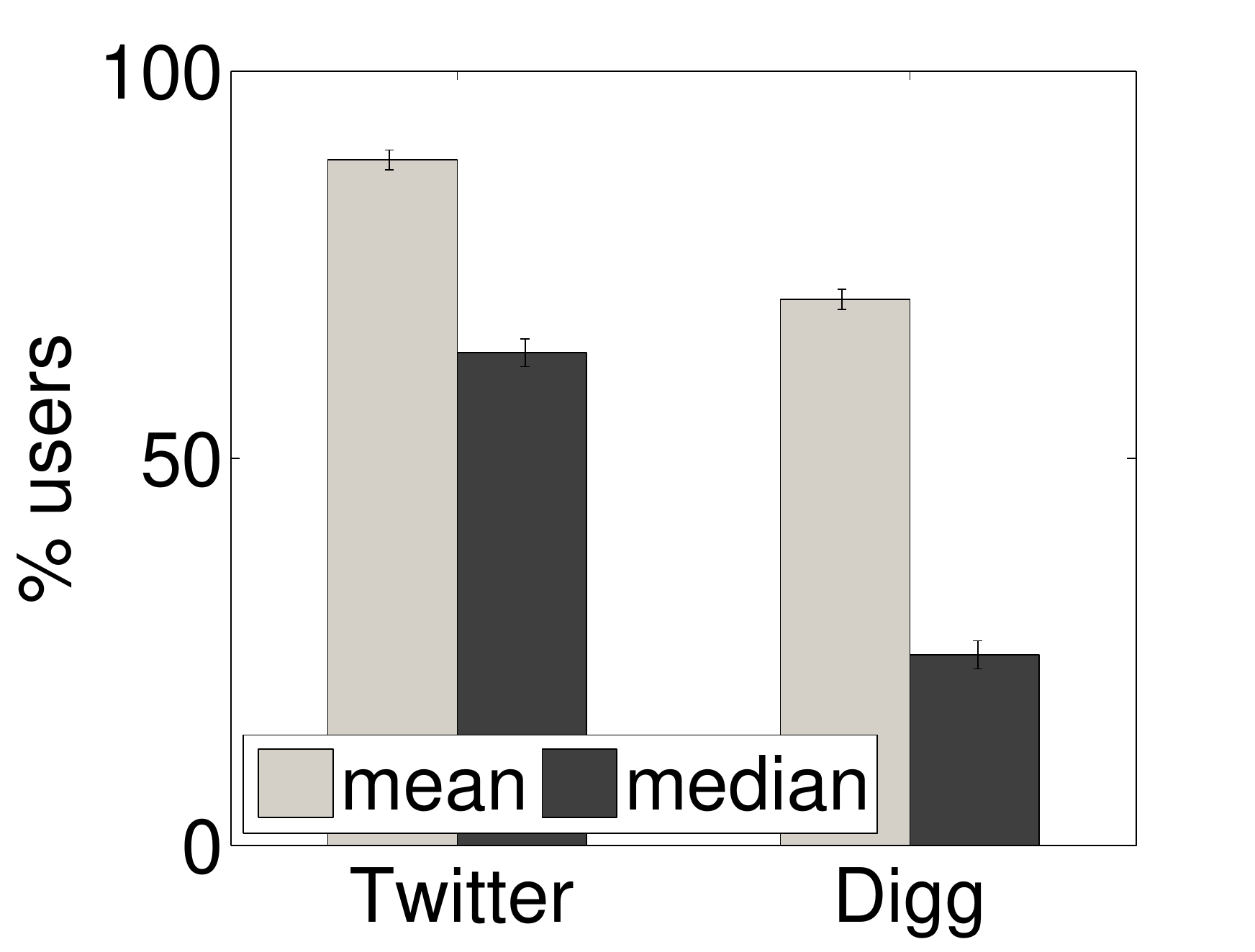}
}
&
\subfigure[Received URL virality]{%
\label{fig:virality_rec_assortativity_shuff}
\includegraphics[width=0.5\columnwidth]{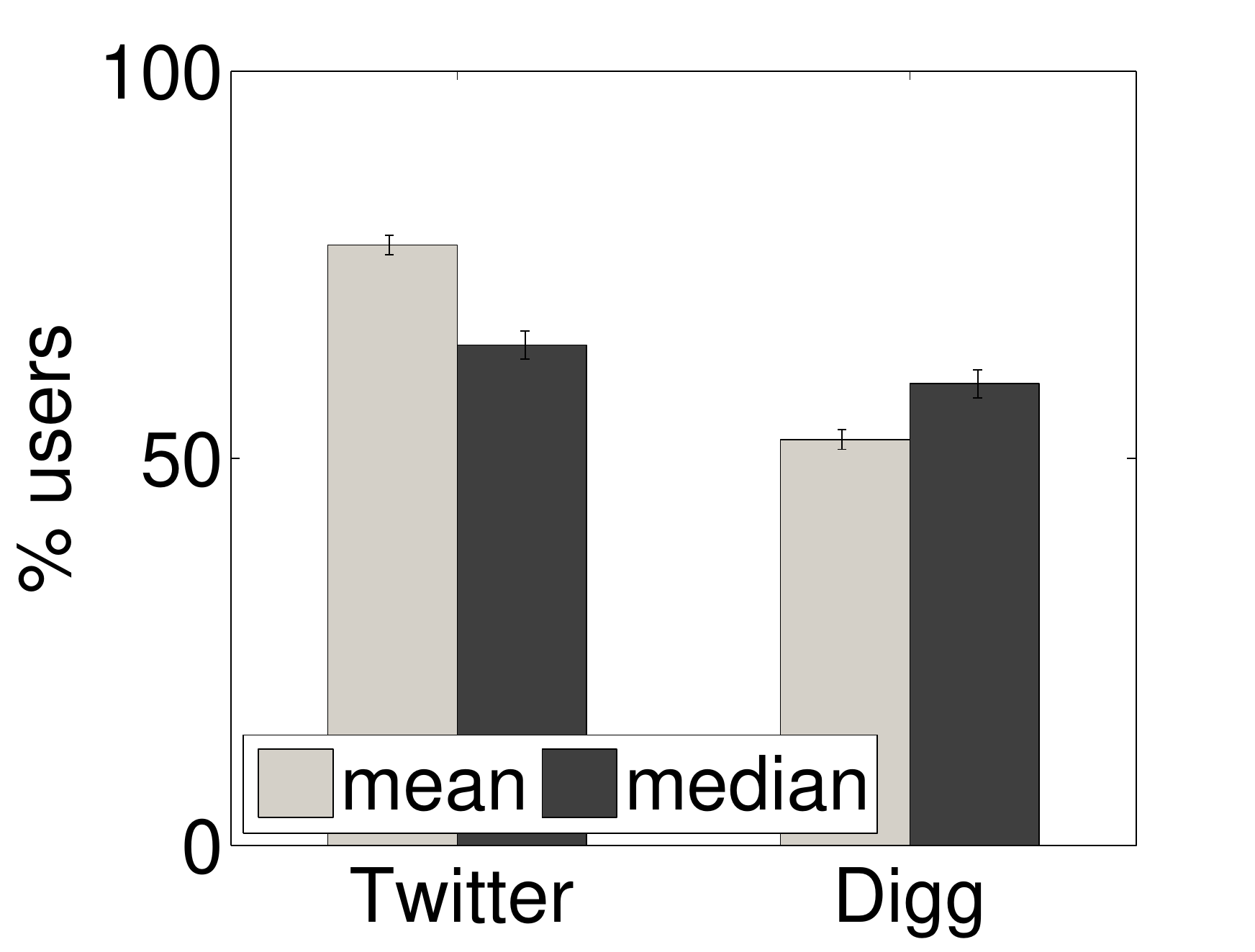}
}
\end{tabular}
\vspace*{-2mm}
\caption{Percentage of users in paradox regime for the shuffled attribute, but keeping
attribute-connectivity correlation (controlled shuffling).
}
\label{fig:attr_attr_shuffle}
\vspace*{-2mm}
\end{center}
\end{figure}

Figure~\ref{fig:attr_attr_shuffle} shows the result of the controlled shuffle test, which should be compared with empirical data in Fig.~\ref{fig:paradoxes}.
No single type of correlation is responsible for all paradoxes.
The activity paradox is greatly reduced by controlled shuffle of the Twitter network and disappears on Digg, suggesting that between-node correlation (here activity assortativity) is largely responsible for this paradox. This means that the paradox arises because active users preferentially link to other active users. The posted-URL virality paradoxes are similar in that it is largely reduced by controlled shuffling. We conclude that it mostly arises due to assortativity of this attribute -- not within-node correlation.
The diversity paradox, however, appears to be unaffected by controlled shuffling both on Digg and Twitter. This suggests that  the diversity paradox is not caused by assortativity of diversity. Instead, it is due to within-node degree--diversity correlation.
The received-URL virality paradox is similar to diversity in that it is largely unaffected by controlled shuffling. Hence, we conclude that degree--virality correlation plays a key role in creating the paradox, but it is not simply due to users selectively seeking out interesting users.

%We observed that the paradoxes arise from certain correlations in users' attributes and degrees. But why do these correlations exist?
There are a few plausible behavioral mechanisms that could lead to these correlations. First, some of the correlations arise simply from the nature of the attribute. For example the within-node correlation of number of friends and diversity (as measured by number of distinct URLs) exists because as users add more friends, they will eventually begin connecting with users outside of their immediate interests.  Between-node correlations arise as users position themselves near people with certain characteristics. For example, active users are generally more engaged with the social networking site, consuming more information. To increase the amount of content they receive, they could seek out new friends (degree--activity correlation) or seek out more active friends (activity assortativity).  Regardless of which correlation may be indicated as the source of the strong paradox for a particular attribute, the unifying theme is that they arise from decisions made by the users and not from statistical artifacts of averaging.

\section{Related Work}\label{sec:relatedwork}
\noindent The friendship paradox and other paradoxes have been shown to occur in a variety of contexts, both online and offline.  As many have pointed out, starting with Feld~\cite{Feld91}, these network paradoxes arise from users with high degree being overrepresented in the population of friends.   Feld also claimed that the friendship paradox, that your friends have higher degree than you, holds with both the mean and the median.  In this paper we build on existing works which have documented paradoxes using the mean, including~\cite{garcia2012using,Christakis10,zuckerman2001makes,Ugander11,Eom14,Hodas13icwsm}.

Hodas et al. demonstrated a variety of paradoxes on Twitter beyond the friendship paradox~\cite{Hodas13icwsm}.  They showed that users' friends and followers are more active and more highly connected. They claimed that any user attribute correlated with connectivity will ultimately result in a paradox.  They don't establish if the activity or virality paradox results simply from the nature of heavy-tailed distributions or from how users choose to position themselves in the network, i.e. behavioral factors.   On the other hand, Hodas et al. identify some key paradoxes that cannot be explained simply by correlations between degree and activity, such as how overloaded users receive more viral content than underloaded users.

Eom and Jo recently examined `generalized friendship paradoxes' in coauthorship networks~\cite{Eom14}.  They showed that correlation between degree and any user characteristic may ultimately lead to a paradox in the mean between the average neighbor and the average user. Although they do not explore the differences between median and mean, and thus do not surprisingly find numerous new paradoxes, they identify some qualities of authors correlated with their degree, including citations per publication and number of publications. In short, they find that your coauthors are more prolific and highly cited than you are.

The present paper takes a closer look at the origin of various paradoxes on social networks and tests how they depend on the statistical methods employed.  We reveal essential differences between utilizing the mean versus the median.  For example, friendship paradoxes on shuffled networks disappear when using the median, revealing it is not simply due to statistical overrepresentation.

\section{Conclusion}

A network paradox exists when one expects the value of some user attribute (degree, activity, etc.) to be less than the average value of this attribute among her neighbors.  Such paradoxes have been observed in many social networks for a variety of attributes. Although there exist explanations for the traditional friendship paradox in undirected networks, in this work we proposed causes of these paradoxes originating in correlations between user attributes the choices users make to correlate themselves with their neighbors. We showed that when the attribute has a heavy-tailed distribution, as is often the case for social attributes, the paradox always exists when mean is used to measure the average, regardless of the underlying system, as long as the mean of the attribute is larger than the median.  However, utilizing the median does not inevitably lead to a paradox. %This stems from statistical properties of heavy-tailed distributions.
%This is because sample mean of values drawn from a heavy-tailed distribution is always greater than sample median.
Surprisingly, most of the paradoxes observed in online social networks still hold when sample median is used. This allows us to restate network paradoxes in their strong form: the \emph{majority} of your friends are better connected, more active, and exposed to more viral and diverse content than you.
%We confirm the stronger network paradoxes for a variety of user attributes on two online social networks of Twitter and Digg.

We probed the behavioral origins of the paradoxes using the shuffle test. This test eliminates correlations between node degrees and attributes by shuffling attribute values of nodes. We show that the friendship paradox with the median disappears on Digg and Twitter after shuffling node degrees, suggesting the correlation of directed node degrees gives rise to the strong friendship paradox on directed networks. Shuffling other user attributes has a similar effect, leading us to conclude that the between-node correlation between attributes of connected nodes and the within-node correlation between user degree and attribute produce network paradoxes for different attributes. Furthermore, we conducted a controlled shuffle test to distinguish the effects of within-node and between-node correlations and we found that for activity and virality of posted content the between-node correlation has a major role, whereas for diversity and virality of received content within-node correlation is the key factor. More research remains to be done to clarify how attribute assortativity is related to within-node correlation, how these correlations affect behavior, and whether they are both caused by some other network factors.

Existence of the strong paradox for an attribute implies that, for most users, a randomly selected friend is likely to exceed the user in that attribute. Because the same behavioral factors that  create between-node correlations in attributes and within-node correlations  are often related to desirability of that attribute (extraversion, wealth, etc.), as people attempt to maximize that attribute, they end up dynamically positioning themselves in the network to remain subject to the strong paradox. 
%The user will perceive themselves as inferior to friends, even if comparison is done on a one-to-one basis. This may explain why self-assessments are negatively correlated with exposure to online social media~\cite{kross2013facebook,chou2012they}.

% new values from the distribution to users, which preserves the distribution. We observe that the paradoxes still strongly exist for the mean, but they do not exist anymore for the median. This is in tune with our theoretical findings. So, we conclude for the behavior different correlations might have verb the cause. To understand the role of each correlation, we perform a controlled shuffle test, which preserve the degree-attribute correlation, but destroys attribute-attribute correlation. We do this by grouping users with the similar number of friends and shuffling the values within each group. The results suggest that the correlations play different roles for different paradoxes. For some paradoxes connectivity-attribute correlation is the key factor, for some others the attribute-attribute, and for the rest both correlations are equally important.

Our findings shed light on the causes of network paradoxes in social networks, and these paradoxes have implications in network sensing, early detection of outbreaks, and users' perceptions of their world. The present work suggests that  models of network formation which cannot reproduce observed paradoxes are not successfully capturing behavioral factors that cause users to correlate themselves with their neighbors. 

\subsection*{Acknowledgements}
This material is based upon work supported in part by  the Air Force Office of Scientific Research under Contract Nos. FA9550-10-1-0569,  by the Air Force Research Laboratories under contract FA8750-12-2-0186, by the National Science Foundation under Grant No. CIF-1217605, and by DARPA under Contract No. W911NF-12-1-0034.

%\vspace*{-2mm}
{% \small
%\bibliographystyle{aaai}
%\bibliography{references}

}

\end{document}